\documentclass[journal]{IEEEtran}
\usepackage[utf8]{inputenc}
\usepackage{cite}
\usepackage{comment}
\usepackage{amsmath,amssymb,amsfonts}
\usepackage{graphicx}
\usepackage{textcomp}
\usepackage[dvipsnames]{xcolor}
\usepackage{footmisc}
\usepackage{url}

\graphicspath{{figs/}{bio-photos/}}

\usepackage[binary-units=true]{siunitx}
\usepackage{xspace}
\usepackage{float}

\newcommand{\ms}{\,\si{\milli\second}\xspace}

\newcommand{\mJ}{\,\si{\milli\joule}\xspace}

\newcommand{\new}[1]{{\color{black}#1}} 

\newcommand{\rebuttal}[1]{{\color{black}#1}} 

\newcommand{\figref}[1]{Fig.~\ref{#1}}

\newcommand{\secref}[1]{Sec.~\ref{#1}}

\usepackage[]{svg} 
\svgpath{02_figures/}

\usepackage{comment}

\usepackage{nohyperref}

\usepackage{xcolor}

\definecolor{matplotlib0}{HTML}{1f77b4}
\definecolor{matplotlib1}{HTML}{d62728}
\definecolor{matplotlib2}{HTML}{2ca02c}
\definecolor{matplotlib3}{HTML}{ff7f0e}
\definecolor{matplotlib4}{HTML}{9467bd}
\definecolor{matplotlib5}{HTML}{8c564b}
\definecolor{matplotlib6}{HTML}{e377c2}
\definecolor{matplotlib7}{HTML}{7f7f7f}
\definecolor{matplotlib8}{HTML}{bcbd22}
\definecolor{matplotlib9}{HTML}{17becf}

\usepackage{amsmath}   
\usepackage{mathtools} 

\DeclarePairedDelimiter\abs{\lvert}{\rvert}%
\DeclarePairedDelimiter\norm{\lVert}{\rVert}%

\usepackage{booktabs}
\usepackage{multirow}
\usepackage{colortbl}
\usepackage{tablefootnote}
\usepackage{threeparttable}
\usepackage{makecell}

\usepackage[acronym, style=super, nonumberlist]{glossaries}

\newcommand{\riscy}[0]{\textsc{RI5CY}}
\newcommand{\eegnet}[0]{EEGNet}

\newcommand{\wolfe}{Mr.\,Wolf\xspace}
\newcommand{\vega}{Vega\xspace}

\usepackage{tikz}
\usetikzlibrary{shapes, arrows}
\tikzset{>=latex}

\usepackage{setspace} 

\tikzstyle{block} = [draw, thick, rectangle, minimum height=0.75cm, minimum width=0.75cm]
\tikzstyle{sum} = [draw, fill=white, circle, node distance=1cm, thick]
\tikzstyle{gain} = [
  regular polygon,
  regular polygon sides=3,
  draw,
  thick,
  fill=white,
  text width=0.3em,
  inner sep=0.3mm,
  outer sep=0mm,
  shape border rotate=-90
]
\tikzstyle{revgain} = [
  regular polygon,
  regular polygon sides=3,
  regular polygon rotate=90,
  draw,
  thick,
  fill=white,
  text width=0.3em,
  inner sep=0.3mm,
  outer sep=0mm,
]

\usepackage{pgfplots}
\usepgfplotslibrary{fillbetween}
\usepgfplotslibrary{colormaps}
\definecolor{color0}{rgb}{0.12156862745098,0.466666666666667,0.705882352941177} 
\definecolor{color1}{rgb}{1,0.498039215686275,0.0549019607843137}
\definecolor{color2}{rgb}{0.172549019607843,0.627450980392157,0.172549019607843} 
\definecolor{color3}{rgb}{0.83921568627451,0.152941176470588,0.156862745098039} 
\definecolor{color4}{rgb}{0.580392156862745,0.403921568627451,0.741176470588235}
\pgfplotsset{
            compat=1.16,
            tick label style={font=\footnotesize},
            legend style={font=\footnotesize}
            }

\pgfplotscreateplotcyclelist{matplotlib}{
  {matplotlib0},
  {matplotlib1},
  {matplotlib2},
  {matplotlib3},
  {matplotlib4},
  {matplotlib5},
  {matplotlib6},
  {matplotlib7},
  {matplotlib8},
  {matplotlib9}
}

\pgfplotsset{every axis/.append style={
    cycle list name=matplotlib,
}}

\usepackage{listings}

\definecolor{code_default}{HTML}{000000}
\definecolor{code_keyword}{HTML}{AC4142}
\definecolor{code_identifier}{HTML}{D28445}

\lstdefinelanguage{RISCV}{
  sensitive=false,
  morecomment=[l]{//},
  alsoletter={.},
  morekeywords=[1]{
    lp.setup, mv, lw, p.lw, sw, p.sw, pv.sdotsp.b, pv.shuffle2.b, p.subNR, p.addNR
  },
  morekeywords=[2]{
    zero, ra, sp, gp, tp, t0, t1, t2, t3, t4, t5, t6, s0, s1, a0, a1, a2, a3, a4, a5, a6, a7, a8, a9, a10, a11,
  },
  morestring=[b]",
  morestring=[b]',
}[strings, comments, keywords]

\lstdefinestyle{RISCV_STYLE}{
  language=RISCV,
  numbers=none,
  basicstyle=\scriptsize\ttfamily\color{code_default},
  keywordstyle=[1]\color{matplotlib0},
  keywordstyle=[2]\color{matplotlib1},
  float,
  captionpos=b,
  belowskip=-0.5cm
}

\usepackage{algorithm}
\usepackage{algpseudocode}
\usepackage{float}
\newfloat{algorithm}{t}{top}

\newacronym{simd}{SIMD}{single instruction, multiple data}
\newacronym{elu}{ELU}{exponential linear unit}
\newacronym{relu}{ReLU}{rectified linear unit}
\newacronym{rpr}{RPR}{random partition relaxation}
\newacronym{mac}{MAC}{multiply-and-accumulate}
\newacronym{dma}{DMA}{direct memory access}
\newacronym{cnn}{CNN}{convolutional neural network}
\newacronym{dnn}{DNN}{deep neural network}
\newacronym{bmi}{BMI}{brain--machine interface}
\newacronym{bci}{BCI}{brain--computer interface}
\newacronym{smr}{SMR}{sensory motor rhythms}
\newacronym{eeg}{EEG}{electroencephalography}
\newacronym{svm}{SVM}{support vector machine}
\newacronym{svd}{SVD}{singular value decomposition}
\newacronym{evd}{EVD}{Eigendecomposition}
\newacronym{iir}{IIR}{infinite impulse response}
\newacronym{fir}{FIR}{finite impulse response}
\newacronym{fc}{FC}{fabric controller}
\newacronym{nn}{NN}{neural network}
\newacronym{mrc}{MRC}{multispectral Riemannian classifier}
\newacronym{flop}{FLOP}{floating point operation}
\newacronym{sos}{SOS}{second-order section}
\newacronym{ipc}{IPC}{instructions per cycle}
\newacronym{tcdm}{TCDM}{tightly coupled data memory}
\newacronym{fpu}{FPU}{floating point unit}
\newacronym{fma}{FMA}{fused multiply-add}
\newacronym{alu}{ALU}{arithmetic logic unit}
\newacronym{dsp}{DSP}{digital signal processing}
\newacronym{gpu}{GPU}{graphics processing unit}
\newacronym{soc}{SoC}{system-on-chip}
\newacronym{mi}{MI}{motor imagery}
\newacronym{csp}{CSP}{common spatial patterns}
\newacronym{fbcsp}{FBCSP}{filter-bank \acrlong{csp}}
\newacronym{pulp}{PULP}{parallel ultra-low power}
\newacronym{soa}{SoA}{state-of-the-art}
\newacronym{bn}{BN}{batch normalization}
\newacronym{isa}{ISA}{instruction set architecture}
\newacronym{mmm}{MMM}{matrix-matrix multiplication}
\newacronym{mcu}{MCU}{microcontroller unit}
\newacronym{ssvep}{SSVEP}{steady-state visual evoked potential}
\newacronym{fpga}{FPGA}{field programmable gate array}
\newacronym{sram}{SRAM}{static random-access memory}
\newacronym{mram}{MRAM}{magnetoresistive random-access memory}
\newacronym{mlp}{MLP}{multi-layer perceptron}

\usepackage{tikz}
\usetikzlibrary{shapes, arrows, snakes}
\tikzset{>=latex}

\usepackage{pgfplots}
\pgfplotsset{compat=1.16}
\definecolor{colorblue}{rgb}{0.12156862745098,0.466666666666667,0.705882352941177} 
\definecolor{colorgreen}{rgb}{0.172549019607843,0.627450980392157,0.172549019607843} 
\definecolor{colorred}{rgb}{0.83921568627451,0.152941176470588,0.156862745098039} 
\pgfplotsset{every axis/.append style={
font=\footnotesize,
label style={font=\footnotesize},
tick label style={font=\scriptsize}  
}
}

\usepackage{filecontents}

\makeatletter
\newcommand\notsotiny{\@setfontsize\notsotiny\@vipt\@viipt}
\makeatother

\usepackage{fancyhdr}
\fancypagestyle{mahmood}{%
  \fancyhf{} 
  
  \fancyhead[C]{\footnotesize \textcopyright 2021 IEEE. Personal use of this material is permitted.  Permission from IEEE must be obtained for all other uses, in any current or future media, including reprinting/republishing this material for advertising or promotional purposes, creating new collective works, for resale or redistribution to servers or lists, or reuse of any copyrighted component of this work in other works.}
}%
\makeatletter
\let\ps@IEEEtitlepagestyle\ps@mahmood
\makeatother

\begin{document}
\bstctlcite{IEEEexample:BSTcontrol}

\title{Sub-100$\mu$W Multispectral Riemannian Classification for EEG-based Brain--Machine Interfaces}

\author{Xiaying~Wang,~\IEEEmembership{Student Member,~IEEE,}
        Lukas~Cavigelli,~\IEEEmembership{Member,~IEEE,}
        Tibor~Schneider,~\IEEEmembership{Student Member,~IEEE,}
        Luca~Benini,~\IEEEmembership{Fellow,~IEEE\vspace{-0.4cm}}
\thanks{Manuscript received August 31, 2021; revised October 30, 2021; accepted December 08, 2020.}
\thanks{X. Wang and L. Benini are with the Integrated Systems Laboratory, Department of Information Technology and Electrical Engineering (D-ITET), ETH Zürich, Switzerland (e-mails: \{xiaywang, benini\}@iis.ee.ethz.ch). L. Cavigelli is with Huawei Technologies, Computing Systems Lab, Zurich Research Center, Switzerland (e-mail: lukas.cavigelli@huawei.com). T. Schneider worked previously at the Integrated Systems Laboratory and is currently with the Networked Systems Group, D-ITET, ETH Zürich, Switzerland (e-mail: sctibor@ethz.ch).} 
\thanks{This project was supported by the Swiss Data Science Center (SDSC) Ph.D. Fellowship under grant ID P18-04.}
}

\maketitle{}

\begin{abstract}
\new{
\acrfull{mi} \acrfullpl{bmi} enable us to control machines by merely thinking of performing a motor action. Practical use cases require a wearable solution where the classification of the brain signals is done locally near the sensor using machine learning models embedded on energy-efficient \acrfullpl{mcu}, for assured privacy, user comfort, and long-term usage. 
In this work, we provide practical insights on the accuracy-cost trade-off for embedded \acrshort{bmi} solutions. Our multispectral Riemannian classifier reaches 75.1\% accuracy on a 4-class \acrshort{mi} task. The accuracy is further improved by tuning different types of classifiers to each subject, achieving 76.4\%. We further scale down the model by quantizing it to mixed-precision representations with a minimal accuracy loss of \rebuttal{1\% and 1.4\%, respectively, which is still up to 4.1\% more accurate than the state-of-the-art embedded convolutional neural network}. We implement the model on a low-power \acrshort{mcu} within an energy budget of merely {198$\mu$J} and taking only {16.9\,ms} per classification. Classifying samples continuously, overlapping the 3.5\,s samples by 50\% to avoid missing user inputs allows for operation at just {85$\mu$W}. Compared to related works in embedded \acrshort{mi}-\acrshortpl{bmi}, our solution sets the new state-of-the-art in terms of accuracy-energy trade-off for near-sensor classification.
}
\end{abstract}

\begin{IEEEkeywords}
brain--machine interface, motor imagery, edge computing, parallel computing, machine learning.
\end{IEEEkeywords}

\IEEEpeerreviewmaketitle{}

\section{Introduction}\label{ch:introduction}

\new{
\IEEEPARstart{B}{rain}--machine interfaces (\acrshortpl{bmi}\glsunset{bmi}) provide a direct communication pathway between the human brain and an external device, such as a wheelchair~\cite{Kobayashi2018BCI-basedEEG-short}, a prosthetic arm~\cite{VILELA2020prosthesis}, or a drone~\cite{Koizumi2018_BMI_visual}. They are especially useful for individuals with physical disabilities to regain independence~\cite{Frolov2017Post-strokeTrial,Kobayashi2018BCI-basedEEG-short}, and are often used in biomedical and clinical scenarios such as stroke rehabilitation~\cite{Biasiucci2018Brain-actuatedStroke}. Particularly interesting is the \gls{mi} \gls{bmi}, where the subject's intention is decoded from the brain activities recorded when the movement of a body part, e.g., feet, is merely imagined. The most common non-invasive technology to acquire brain signals is based on \gls{eeg} thanks to its portability and relatively {inexpensive hardware setup.} A growing number of commercial {\gls{eeg} devices have been recently announced, making} \glspl{bmi} increasingly common also outside laboratory and clinical environments~\cite{Valentin2019realworld,marin2020emotiv}.

With the rising popularity of wearable \gls{bmi} devices, it is crucial for a safe and successful \gls{bmi} to meet key requirements, such as accurate {classification}, user comfort, long-term usage, real-time response, and, last but not least, privacy preservation. Traditional \gls{bmi} systems collect and send the user data to a connected computer, a cloud, or a gateway with a wired or wireless setup, and often adopt offline, remote data processing to extract useful information.
Connected cables yield a bulky \gls{bmi} setup limiting subjects' degrees of movement freedom. While wireless technology allows better user comfort, it introduces major drawbacks such as short battery life and privacy concerns~\cite{beach2021edgealgorithmswearables}. Moreover, offline analysis leads to long delays in the information extraction.
Recently, academic and industrial researchers have shared increasing interest in the concept of edge computing where the data is not remotely transmitted but directly processed near the sensor node~\cite{zhu2020tbiocas,wong2021edge}. 
The collected sensitive biomedical signals are directly processed on-the-fly near the sensor right where the data is acquired and only the outcome of the processing is delivered, effectively {curtailing communication-induced latency and} eliminating most security and privacy risks introduced by the transmission and a remote processing pipeline.

However, many challenges need to be addressed {in edge processing for BMI devices}. 
Analyzing a large amount of sensor data at the edge requires miniaturized processing engines that provide enough compute power to extract useful information in real-time, while at the same time consume low power and are energy-efficient for a prolonged battery life~\cite{beach2021edgealgorithmswearables}. A popular family of \glspl{mcu} is the ARM Cortex-M series. They are commercialized {in} a wide variety ranging from the small and very-low power Cortex-M0 to the fast and {high-performance} Cortex-M7 devices. Another {recently introduced edge-processing approach} is the \gls{pulp} platform based on the open-source RISC-V \gls{isa}~\cite{Flamand2018gap8}. Previous studies have proven that \gls{pulp} processors outperform the Cortex-M series in terms of low power consumption, energy efficiency, and high compute capabilities~\cite{wang2019fann}, also for \gls{mi}-\gls{bmi} applications~\cite{Schneider2020}. Moreover, the \wolfe processor~\cite{pullini2019wolf} of the \gls{pulp} family has been deployed on a miniaturized \gls{bmi} system, called Biowolf~\cite{kartsch2019biowolf}. Thanks to the eight low-power, parallel processing units of \wolfe, the Biowolf platform is one of the {leading-edge} systems for \gls{bmi} applications.
A next-generation \gls{pulp} processor has been very recently released, called \vega~\cite{Rossi2021vega}. Compared to \wolfe, it provides more performance and better energy efficiency with its nine parallel cores, four \glspl{fpu}, and more on-chip memory. No work has yet demonstrated its capabilities with a real application.

}

\new{
Resource-efficient yet accurate classification algorithms are also crucial for a successful smart wearable device~\cite{beach2021edgealgorithmswearables}, especially for \gls{eeg} applications where} the high variability across subjects and among different recording sessions poses big challenges~\cite{zhu2020tbiocas}.
\new{\Glspl{dnn} have demonstrated impressive results in many fields, from outperforming humans in computer vision \cite{russakovsky2015imagenet}, vastly improving solutions for image processing \cite{cavigelli2017cas} and speech recognition \cite{deng2013new}, natural language processing \cite{8416973}, large-scale recommender systems \cite{9216015}, and data analysis for sensors such as radar \cite{seyfiouglu2018deep} that are not directly understandable by humans. 
Especially \glspl{cnn} are largely applied in the image domain, but also for \gls{mi}-\gls{bmi} classification achieving \gls{soa} accuracy~\cite{Schirrmeister2017DeepVisualization,Lawhern2018EEGNet:Interfaces,wu2019_MSFBCNN,ingolfsson2020eegtcnet}. However, they tend to grow in numbers of parameters making them often unfit for deployment on low-power low-cost \glspl{mcu}, and require a large amount of training data to prevent overfitting. Whereas for \gls{bmi} applications the acquisition and the} labeling of \gls{eeg} data are expensive, time-consuming, and prone to errors, resulting in scarce amounts of data available for training complex models with large numbers of parameters~\cite{leon2020}.
Over the years, successful methods have been proposed to extract discriminative, domain-specific features from \gls{eeg} signals. The well-known \gls{csp} algorithm learns spatial filters that discern between different \gls{mi} tasks~\cite{lotte2018_10yearreview}. An improved algorithm, called \gls{fbcsp}, that accounts for multiple frequency bands has achieved better accuracy~\cite{KaiKengAng2008FilterInterface,park2018fbcsp}. More recent studies have proposed Riemannian methods to extract more comprehensive features also in absence of labeled data~\cite{NGUYEN20181871,wu2021riemannian_recent}.
The unsupervised feature calibration enables online adaptation of the classifier to combat the large inter-session variance in MI-BMIs~\cite{lotte2019Adaptive}. 
So far, these methods are believed to be the most promising feature extractors for several kinds of \gls{bmi} paradigms~\cite{congedo2017RiemannianReview,Yger2017RiemannianReview,riemannianSSVEPreview}. 

\new{
Few studies can be found in literature that have deployed \gls{mi}-\gls{bmi} models on edge devices. Bewalfi et al.~\cite{bewalfi2018_wolacsp} and Malekmohammadi et al.~\cite{malekmohammadi2019_cspsvm} have proposed \gls{csp} solutions based on \glspl{fpga}. While Schneider et al.~\cite{Schneider2020} have deployed a compact \gls{cnn} on \wolfe achieving an accuracy of 71\% on the popular BCI Competition IV-2a dataset and an energy consumption of 0.34\mJ, making it the embedded \gls{bmi} with the lowest energy utilization. Wang et al.~\cite{wang2021mrc}, on the other hand, have proposed an implementation of \gls{mrc} on the same platform achieving 3\% better accuracy while consuming 1.3\mJ, showing the trade-off between accuracy and cost.

In this work, we propose {and release open-source}\footnote{\url{https://github.com/pulp-platform/multispectral-riemannian}\label{footnote:code}} an improved implementation of \gls{mrc} in terms of both accuracy and cost on the novel parallel platform \vega~\cite{Rossi2021vega}. The main contributions are summarized as follows:
\begin{itemize}
    \item \rebuttal{We achieve an accuracy of 75.1\% on a 4-class \gls{mi} task when using only linear \glspl{svm}. The performance is further improved to 76.4\% by tailoring the classifier to the subject within a given resource constraint, i.e., by tuning the hyperparameters of different types of classifiers, namely \glspl{svm} and \glspl{mlp}, for each subject.} 
    \item We quantize the feature extraction and the classifier of the \gls{mrc} from full 32-bit float precision to a combination of \mbox{8-}, \mbox{16-}, \mbox{32-bit} \mbox{fixed-} and floating-point representations, to maximize the efficiency on the hardware, while at the same time preserving similar accuracy. The quantization yields a 1.4\% accuracy loss which is still 0.9\% more accurate than the previous implementation of \gls{mrc}~\cite{wang2021mrc} and 4.1\% more accurate than the \gls{soa} embedded \gls{cnn} with the lowest energy consumption~\cite{Schneider2020}.
    \item We implement for the first time a real application on the new \vega platform by demonstrating its potentials for low-power near-sensor analytics. 
    We measure the runtime and power consumption on-board and find the configurations with the highest energy efficiency and the shortest execution time. Experimental measurements show that the former yields {16.9\,ms} and only {198\,\textmu J} and the latter takes only {7.9\,ms} and consumes {338\,\textmu J} per inference. Together with the high classification accuracy, it becomes the new \gls{soa} embedded \gls{bmi} with the highest classification accuracy and the lowest energy utilization.
\end{itemize}

}

\section{Related Work}\label{ch:relworks}

\begin{figure}
    \centering

  \resizebox{\columnwidth}{!}{%

\begin{tikzpicture}

\begin{axis}[
name=myAxis,
width=\columnwidth, height=0.75\columnwidth,
legend cell align={left},
legend columns=1,
legend style={
  at={(0.99,0.5)},
  anchor=east,
  draw=none,
  font=\scriptsize
},
log basis x={10},
tick align=inside,
tick pos=left,
x grid style={black!20},
xlabel={Memory footprint [kB]},
xmajorgrids,
xmin=10, xmax=507847,
xmode=log,
xlabel near ticks,
y grid style={black!20},
ylabel={Accuracy [\%]},
ymajorgrids,
ymin=64, ymax=80,
ytick={64, 66, ..., 80},
yticklabels={64, 66, ..., 80},
ylabel near ticks,
clip=false
]



\addplot[thick, draw=white!69.0196078431373!black, dashed, forget plot] coordinates {(64,64)(64,80)};
\node[] at (axis cs: 50,80.4) {\notsotiny 64\,kB};
\addplot[thick, draw=white!69.0196078431373!black, dashed, forget plot] coordinates {(512,64)(512,80)};
\node[] at (axis cs: 512,80.4) {\notsotiny 512\,kB};
\addplot[thick, draw=white!69.0196078431373!black, dash dot, forget plot] coordinates {(128,64)(128,80)};
\node[] at (axis cs: 150,80.4) {\notsotiny 128\,kB};
\addplot[thick, draw=white!69.0196078431373!black, dash dot, forget plot] coordinates {(1500,64)(1500,80)};
\node[] at (axis cs: 2000,80.4) {\notsotiny 1500\,kB};

\fill[white!90!black] (1000,66.8) circle (0.1025cm) node[above=0.1025cm, yshift=-0.5ex, black] {\scriptsize 1\,M};
\fill[white!90!black] (3500,66.8) circle (0.225cm) node[above=0.225cm, yshift=-0.5ex, black] {\scriptsize 50\,M};
\fill[white!90!black] (16000,66.8) circle (0.35cm) node[above=0.35cm, yshift=-0.5ex, black] {\scriptsize 100\,M};
\fill[white!90!black] (130000,66.8) circle (0.6cm) node[above=0.6cm, yshift=-0.5ex, black] {\scriptsize 200\,M};

\fill[colorred] (68.15,70.9) circle (0.1325cm) node[below=0.1325cm, xshift=-1.5ex, yshift=-0.5ex, black] {\notsotiny Q-EEGNet~\cite{Schneider2020}};
\addplot[thick, gray, only marks, mark=*, mark size=0.6] coordinates {(68.15,70.9)(68.15,70.9)};
\fill[colorred] (910.192,71.2) circle (0.1325cm) node[below=0.1325cm, xshift=0.5ex, yshift=0.5ex, black] {\notsotiny EEGNet~\cite{Lawhern2018EEGNet:Interfaces}};
\addplot[thick, gray, only marks, mark=*, mark size=0.6] coordinates {(910.192,71.2)(910.192,71.2)};
\fill[colorred] (4241.2,74.31) circle (0.2575cm) node[below=0.2575cm, xshift=2ex, yshift=0.5ex, black] {\notsotiny S. ConvNet~\cite{Schirrmeister2017DeepVisualization}};
\addplot[thick, gray, only marks, mark=*, mark size=0.6] coordinates {(4241.2,74.31)(4241.2,74.31)};
\fill[colorred] (23720,75.80) circle (0.605cm) node[above=0.605cm, xshift=1ex, yshift=-0.5ex, black] {\notsotiny MSFBCNN~\cite{wu2019_MSFBCNN}};
\addplot[thick, gray, only marks, mark=*, mark size=0.6] coordinates {(23720,75.80)(23720,75.80)};
\fill[colorred] (1601.08,77.34) circle (0.117cm) node[above=0.117cm, yshift=-0.5ex, black] {\notsotiny EEG-TCNet~\cite{ingolfsson2020eegtcnet}};
\addplot[thick, gray, only marks, mark=*, mark size=0.6] coordinates {(1601.08,77.34)(1601.08,77.34)};

\fill[colorred] (1244.0,73.70) circle (0.36cm) node[below=0.36cm, xshift=-0ex, yshift=0.5ex, black] {\notsotiny FBCSP~\cite{Hersche2018FastFeatures}};
\addplot[thick, gray, only marks, mark=*, mark size=0.6] coordinates {(1244.0,73.70)(1244.0,73.70)};
\fill[colorgreen] (74,73.82) rectangle (96,74.38) node[below, xshift=-0.5ex, yshift=-0.5ex, black] {\notsotiny \color{colorgreen} MRC-Mr.\,Wolf~\cite{wang2021mrc}};
\addplot[thick, gray, only marks, mark=*, mark size=0.6] coordinates {(84,74.1)(84,74.1)};

\addplot [thick, colorblue]
table {%
mrc-vega.dat
};
\node[] at (axis cs: 250,75.2) {\notsotiny \color{colorblue} MRC-Vega};

    \end{axis}
    
\end{tikzpicture}
}

    \caption{Accuracy vs. memory footprint and computational complexity in terms of \acrshort{mac} operations of related work as well as the trade-off by our method (blue line) and the configuration used for most of our analyses (green). 
    The memory footprint is estimated as the sum of the model size and the biggest consecutive feature maps based on the common layer-by-layer computation paradigm.
    As our method is not dominated by \acrshort{mac} operations as opposed to the competing \acrshort{dnn}-based methods, we discuss execution time separately.}
    \label{fig:acc_mem_comp}
\end{figure}
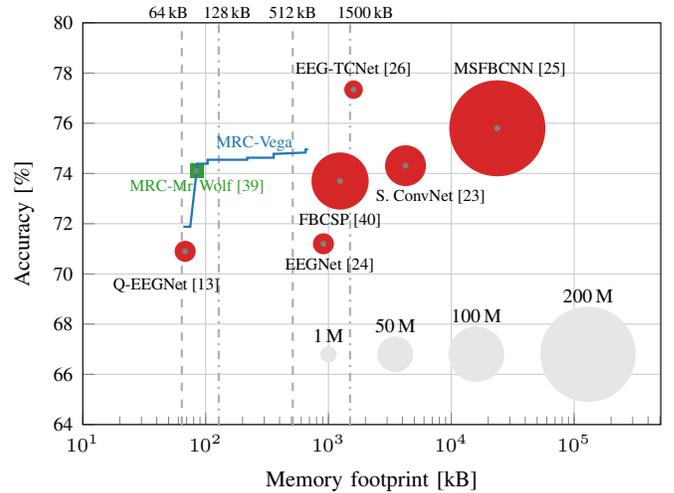

\setlength{\tabcolsep}{3.5pt}
\begin{table}[t]
\caption{\rebuttal{Comparison to related works on BCI Competition IV-2a dataset in terms of accuracy, model size (top), quantized accuracy, computation runtime and energy per inference (bottom).}}
\centering
\label{tab:relworks_implperf}
{\small
  \begin{threeparttable}
\begin{tabular}{@{}lrrrr@{}}
\toprule
 & \multicolumn{2}{c}{Acc. [\%]} & \multicolumn{2}{c}{\multirow{2}[3]{*}{Num. param.}} \\
\cmidrule(l){2-3}
  & \multicolumn{1}{r}{2-}  & \multicolumn{1}{r}{4-class}  &  &  \\ 
\midrule
S. ConvNet\,'17~\cite{Schirrmeister2017DeepVisualization}$^{**}$ & 79.90 & 74.31 & \multicolumn{2}{c}{47.3\,k} \\
EEGNet\,'18~\cite{Lawhern2018EEGNet:Interfaces,Schneider2020} & - & 71.30 & \multicolumn{2}{c}{\textit{2.63\,k}} \\
\acrshort{fbcsp}\,'18~\cite{Hersche2018FastFeatures} & - & 73.7 & \multicolumn{2}{c}{261\,k} \\
\acrshort{mrc}\,'18~\cite{Hersche2018FastFeatures} & - & 75.47 & \multicolumn{2}{c}{32.6\,k} \\
MSFBCNN\,'19~\cite{wu2019_MSFBCNN} & - & 75.8 & \multicolumn{2}{c}{155\,k} \\
EEG-TCNet\,'20~\cite{ingolfsson2020eegtcnet}$^\dagger$ & - & 77.35 & \multicolumn{2}{c}{4.27\,k} \\
MRC-MLP\,'20~\cite{Yang2020ieeeaccess} & - & 76.1 & \multicolumn{2}{c}{5.7\,M} \\
\acrshort{csp}+Riemannian\,'20~\cite{chen2020cssimilarity}$^\diamond$ & - & \textit{77.82} & \multicolumn{2}{c}{-} \\
JSTFD+LDA\,'20~\cite{jiang2020_JSTFD_lda} & 79.6 & - & \multicolumn{2}{c}{-} \\
FB+3D-CNN\,'21~\cite{bang2021fb3dcnn} & \textit{86.96} & - & \multicolumn{2}{c}{46\,M} \\
TS-SEFFNet\,'21~\cite{li2021_TS-SEFFNet_squeezeExcitation} & - & 74.71 &  \multicolumn{2}{c}{282\,k} \\
MRC- \\
\hspace{1mm} -Mr.Wolf\,'21~\cite{wang2021mrc} & - & 75.1 & \multicolumn{2}{c}{4.55\,k} \\
\hspace{1mm} \textbf{-Vega}$^\ddagger$ (ours) & - & \textbf{75.1} & \multicolumn{2}{c}{\textbf{4.55\,k}} \\
& \multicolumn{2}{c}{Quant. acc. [\%]} & Time\,[ms] & En.\,[\textmu J] \\
\cmidrule(r){1-1} \cmidrule(l){2-3} \cmidrule(l){4-4} \cmidrule(l){5-5}
WOLA-CSP\,'18~\cite{bewalfi2018_wolacsp} & 78.85 & - & 430 & 301000 \\
\acrshort{csp}+\acrshort{svm}\,'19~\cite{malekmohammadi2019_cspsvm} & 80.55 & 67.21 & 11.66 & 978 \\ 
Q-EEGNet\,'20\,(50$^\natural$)~\cite{Schneider2020} & - & 70.90 & 28.67 & 337 \\
Q-EEGNet\,'20\,(350$^\natural$)~\cite{Schneider2020} & - & 70.90 & 13.64 & 678 \\
MRC- \\
\hspace{1mm} -Mr.Wolf\,'21\,(100$^{\natural}$)~\cite{wang2021mrc} & - & 74.1 & 33.39 & 1304 \\ 
\hspace{1mm} \textbf{-Vega}\,(350$^{\natural}$) (ours) & - & \textbf{74.1} & \textbf{\textit{7.9}} & \textbf{338} \\ 
\hspace{1mm} \textbf{-Vega}\,(50/160$^{\ddagger}$) (ours) & - & \textbf{74.1} & \textbf{16.9} & \textbf{\textit{198}} \\ 
\bottomrule
\end{tabular}
\begin{tablenotes}\footnotesize
\item $^{**}$ Respectively reproduced in~\cite{bang2021fb3dcnn} and \cite{ingolfsson2020eegtcnet} for 2- and 4-class tasks.
\item $^\dagger$ Improves to \textit{83.84\%} with subject-specific variable network architectures and hyperparameters.
\item $^\diamond$ With subject-specific feature selection based on \gls{eeg} channels.
\item $^\ddagger$ Improves to 76.4\% with subject-specific variable classifier. 
\item $^\natural$ Clock frequencies in MHz. All cores run at the same frequency.
    \item $^\ddagger$ The \acrshort{fc} core runs at 50\,MHz. The cluster cores run at 160\,MHz.
\end{tablenotes}
\end{threeparttable}
}
\end{table}

\new{
The next generation of smart edge \glspl{bmi} have to {manage} the three-way trade-off among (a) algorithmic performance in terms of high classification accuracy, (b) cost in terms of low computational runtime and memory footprint, and (c) power and energy consumption~\cite{beach2021edgealgorithmswearables}.
\rebuttal{Many research works have been conducted to boost the performance of \gls{bmi} algorithms in terms of classification accuracy, while only recently increasing attention has been paid to the resources that are required by these algorithms (see Table~\ref{tab:relworks_implperf}).} Considering the popular 4-class BCI Competition IV-2a dataset~\cite{Brunner2008BCIA}, the reproducible \gls{soa} accuracy is around 77\% reached by EEG-TCNet~\cite{ingolfsson2020eegtcnet}, when no additional algorithmic optimizations are used, such as subject-specific feature extraction by tuning the model architecture or subject-specific channel selection. \rebuttal{Compared to other existing works on the same dataset~\cite{Schirrmeister2017DeepVisualization,wu2019_MSFBCNN,Hersche2018FastFeatures,bang2021fb3dcnn,li2021_TS-SEFFNet_squeezeExcitation}, EEG-TCNet requires orders of magnitude fewer resources in terms of memory footprint and computational complexity, while reaching higher classification accuracy, as depicted in Fig.~\ref{fig:acc_mem_comp}.} However, if we consider the common layer-by-layer computation paradigm for the classification inference, the required memory footprint is out-of-reach for the fast on-chip memory of ultra-low power \glspl{mcu} such as \wolfe with 512\,kB L2 and 64\,kB L1 memory.

Bewalfi et al.~\cite{bewalfi2018_wolacsp} are the first ones to propose an embedded \gls{mi}-\gls{bmi} based on \gls{csp} algorithms, called WOLA-CSP, on a Stratix-IV \gls{fpga}. The achieved accuracy is 78.85\% on a 2-class task, while the average power consumption is 0.7\,W with an execution runtime of 430\,ms, yielding an energy consumption of 301\mJ. Malekmohammadi~\cite{malekmohammadi2019_cspsvm} have proposed a much more energy efficient solution based on \gls{csp} and \gls{svm} on a Virtex-6 \gls{fpga}. The 2-class and 4-class classification accuracy scores are respectively 80.55\% and 67.21\%, with an energy consumption of 0.978\mJ.
EEGNet~\cite{Lawhern2018EEGNet:Interfaces}, on the other hand, is a very compact \gls{cnn} with only a few thousand model parameters.}
It has been successfully quantized with \mbox{Q-EEGNet}~\cite{Schneider2020} and implemented on \wolfe ~\new{reaching an energy consumption of only 0.678\mJ.} It has proven to be three orders of magnitude more energy efficient than a EEGNet implementation on commercially available \glspl{mcu} based on ARM \mbox{Cortex-M} architecture~\cite{Wang2020Memea}, making it the \gls{soa} embedded \gls{cnn} in terms of energy efficiency and compact model size, \rebuttal{as reported in Table~\ref{tab:relworks_implperf}.}
\new{However, the classification performance is around 6\% less accurate than the EEG-TCNet~\cite{ingolfsson2020eegtcnet}. Another very promising algorithm to extract discriminant features from \gls{mi} \gls{eeg} data is based on Riemannian geometry~\cite{Yger2017RiemannianReview,Yang2020ieeeaccess}. Hersche et al.~\cite{Hersche2018FastFeatures} have proposed a multiscale temporal and spectral Riemannian classifier with \gls{svm} and achieved a 4-class accuracy of 75.1\% outperforming both EEGNet and \gls{csp}-based models by around 3\%--7\%. It has been further optimized in terms of resource usage by using only one temporal window, quantized to mixed-precision representations, and efficiently implemented on \wolfe with an energy consumption of 1.304\mJ and a runtime of 33.39\ms~\cite{wang2021mrc}. It is the most accurate embedded solution, however, it comes at the cost of a higher energy consumption and a longer runtime compared to~\cite{malekmohammadi2019_cspsvm,Schneider2020}.

In this work, we further improve the classification accuracy of \gls{mrc} and implement it on a more energy-efficient \gls{mcu} called \vega. Similar to \wolfe, \vega is a parallel multi-core processor based on the open-source \mbox{RISC-V} \gls{isa} with custom extensions, but it has an improved hardware architecture and more resources, yielding 3.23$\times$ and 4.39$\times$ better efficiency compared to \wolfe, respectively for integer and floating-point operations~\cite{Rossi2021vega}.}

\begin{figure*}[t]
  \centering
  \begin{tikzpicture}[scale=0.8, transform shape, node distance=3.25cm]
    \node (input) {$\mathbf{X}$};
    \node[block, right of=input, yshift=+0.8cm, xshift=-1cm] (iir1) {IIR $b_1$};
    \node[right of=input, yshift=0.1cm, xshift=-1cm] {$\vdots$};
    \node[block, right of=input, yshift=-0.8cm, xshift=-1cm] (iir2) {IIR $b_n$};
    \node[block, right of=iir1, xshift=-0.4cm] (cm1) {$\hat{\mathbf{X}}_1 \hat{\mathbf{X}}_1^T + \mathbf{I}\rho$};
    \node[right of=iir1, yshift=-0.7cm, xshift=-0.4cm] {$\vdots$};
    \node[block, right of=iir2, xshift=-0.4cm] (cm2) {$\hat{\mathbf{X}}_n \hat{\mathbf{X}}_n^T + \mathbf{I}\rho$};
    \node[block, right of=cm1, xshift=0.2cm] (wh1) {$\mathbf{C}_{\text{ref},1}^{-1/2} \mathbf{C} \mathbf{C}_{\text{ref},1}^{-1/2}$};
    \node[right of=cm1, yshift=-0.7cm, xshift=0.2cm] {$\vdots$};
    \node[block, right of=cm2, xshift=0.2cm] (wh2) {$\mathbf{C}_{\text{ref},f}^{-1/2} \mathbf{C}_n \mathbf{C}_{\text{ref},f}^{-1/2}$};
    \node[block, right of=wh1, xshift=0.2cm] (logm1) {$\text{logm}(\mathbf{W}_1)$};
    \node[right of=wh1, yshift=-0.7cm, xshift=0.2cm] {$\vdots$};
    \node[block, right of=wh2, xshift=0.2cm] (logm2) {$\text{logm}(\mathbf{W}_n)$};
    \node[block, right of=logm1, xshift=-0.4cm] (diag1) {$\text{vect}(\mathbf{L}_1)$};
    \node[right of=logm1, yshift=-0.7cm, xshift=-0.4cm] {$\vdots$};
    \node[block, right of=logm2, xshift=-0.4cm] (diag2) {$\text{vect}(\mathbf{L}_n)$};
    \node[right of=diag1, xshift=-1.75cm, yshift=-0.8cm, rotate=90] (concat) {$\oplus$};
    \node[block, right of=concat, xshift=-2.3cm, yshift=0, rotate=90] (svm) {Classifier (CL)};

    \draw (iir1) node[above, yshift=0.5cm, align=center] {\large Filter};
    \draw (cm1) node[above, yshift=0.5cm, text width=2cm, align=center] {\large Covariance Matrix};
    \draw (wh1) node[above, yshift=0.5cm, align=center] {\large Whitening};
    \draw (logm1) node[above, yshift=0.5cm, text width=2cm, align=center] {\large Matrix Logarithm};

    \draw[thick] (18.20, 0.75) circle (0.1cm);
    \draw[thick] (18.20, 0.25) circle (0.1cm);
    \draw[thick] (18.20, -0.25) circle (0.1cm);
    \draw[thick] (18.20, -0.75) circle (0.1cm);
    \draw[thick] (18.60, -0.75) node[right, rotate=90] {4 classes};

    \draw[->] (input) -- ++(0.75, 0) |- (iir1);
    \draw[->] (input) -- ++(0.75, 0) |- (iir2);
    \path[->] (iir1) edge node[above] {$\hat{\mathbf{X}}_1$} (cm1);
    \path[->] (iir2) edge node[above] {$\hat{\mathbf{X}}_n$} (cm2);
    \path[->] (cm1) edge node[above] {$\mathbf{C}_1$} (wh1);
    \path[->] (cm2) edge node[above] {$\mathbf{C}_n$} (wh2);
    \path[->] (wh1) edge node[above] {$\mathbf{W}_1$} (logm1);
    \path[->] (wh2) edge node[above] {$\mathbf{W}_n$} (logm2);
    \path[->] (logm1) edge node[above] {$\mathbf{L}_1$} (diag1);
    \path[->] (logm2) edge node[above] {$\mathbf{L}_n$} (diag2);
    \draw[->] (diag1) -| (concat) node [above, xshift=-0.3cm, yshift=0.8cm] {$\mathbf{v}_1$};
    \draw[->] (diag2) -| (concat) node [above, xshift=-0.3cm, yshift=-0.8cm] {$\mathbf{v}_n$};
    \draw[->] (concat) -- (svm);
    \path[->] (17.75, 0.75) edge (18.1, 0.75);
    \path[->] (17.75, 0.25) edge (18.1, 0.25);
    \path[->] (17.75, -0.25) edge (18.1, -0.25);
    \path[->] (17.75, -0.75) edge (18.1, -0.75);
    
    \node (rect) at (5.65, 0) [draw,thick,minimum width=8.75cm,minimum height=2.6cm, dashed, color0] {};
    \node (rect) at (5.6, 0.8) [draw=none,thick,minimum width=8.5cm,minimum height=0.85cm, fill=color0, opacity=0.15] {};
    
    \node (rect) at (13.7, 0) [draw,thick,minimum width=5.75cm,minimum height=2.6cm, dashed, color3] {};
    \node (rect) at (13.75, 0.6) [draw=none,thick,minimum width=5.7cm,minimum height=1.25cm, fill=color3, opacity=0.15] {};
    
    \node (rect) at (17.6, 0) [draw,thick,minimum width=1.7cm,minimum height=2.6cm, dashed, color2] {};
  \end{tikzpicture}
  \caption{\acrlong{mrc} with $n=18$ frequency bands and one time window. {The regions surrounded by dashed lines represent the three parts of the computation. The shaded areas represent regions which are processed by all 9 cores in parallel.} Adapted from~\cite{wang2021mrc}.}\label{fig:background:riemannian}
  \vspace{-0.3cm}
\end{figure*}
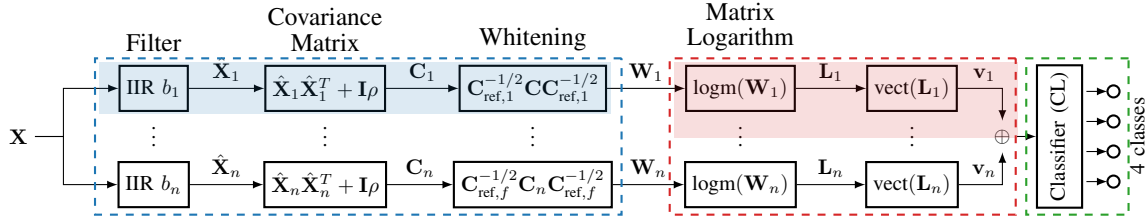

\begin{figure*}[t]
  \centering
  \begin{tikzpicture}[scale=0.78, transform shape, node distance=2.25cm]
    \node (input) {${\mathbf{X}}$};
    \node[block, right of=input, xshift=-0.2cm](iir){$\text{IIR}_{Q12, Q16}$};
    \node[block, right of=iir, xshift=0.5cm](covmat){Covmat$_{Q16}$};
    \node[block, right of=covmat, xshift=0.8cm](white){Whitening$_{Q11}$};
    \node[block, right of=white, xshift=0.8cm](deq){Dequantize};
    \node[block, right of=deq, xshift=0.4cm](logm){$\text{logm}_{F32}$};
    \node[block, right of=logm, xshift=0.4cm](quant){Requantize};
    \node[block, right of=quant, xshift=0.2cm](vec){vect$_{Q8}$};
    \node[block, right of=vec](svm){CL$_{Q8}$};
    \node[draw, circle, thick, right of=svm, xshift=-0.5cm, inner sep=0.065cm] (out) {};

    \draw[->] (input) -- node[above]{Q8} (iir);
    \draw[->] (iir) -- node[above]{Q8} (covmat);
    \draw[->] (covmat) -- node[above]{Q16} (white);
    \draw[->] (white) -- node[above]{Q32} (deq);
    \draw[->] (deq) -- node[above]{F32} (logm);
    \draw[->] (logm) -- node[above]{F32} (quant);
    \draw[->] (quant) -- node[above]{Q8} (vec);
    \draw[->] (vec) -- node[above]{Q8} (svm);
    \draw[->] (svm) -- node[above]{Q32} (out);
  \end{tikzpicture}
  \caption{{Mixed-precision,} quantized \gls{mrc} of a single frequency band, showing the representation of each intermediate signal~\cite{wang2021mrc}.}\label{fig:riemann:feature_extraction_quant}
  \vspace{-0.3cm}
\end{figure*}
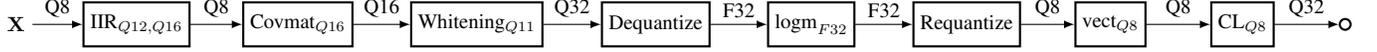

\section{Model Design}\label{ch:methods:model}

\new{

The \gls{mrc} proposed in this work is based on~\cite{wang2021mrc}. Non-linear features are first extracted from different frequency bands using the Riemannian covariance method~\cite{Yger2017RiemannianReview}.
Subsequently, a machine learning classifier is trained to classify different classes. 

\subsection{Feature Extraction}
Fig.~\ref{fig:background:riemannian} illustrates the proposed model. The input \gls{eeg} data $\mathbf{X}$ has dimensions \rebuttal{number of \gls{eeg} channels $N_{ch} =22$ times number of time samples $N_s=875$ and is filtered with $n=18$ \gls{iir} bandpass filters} to obtain the filtered signals $\hat{\mathbf{X}}_k$:
\begin{equation}\label{eq:riemann:design:iir}
  \hat{\mathbf{X}}_k = \text{IIR } b_k(\mathbf{X}),
\end{equation}
with $k \in \{ 1, 2, \ldots, n \}$. 
\rebuttal{They are realized as 4th order Butterworth filters with the frequency bands starting from 4\,Hz until 40\,Hz with a bandwidth of 2\,Hz each, i.e, the lower and upper cut-off frequencies of the \gls{iir} filters are 4\,Hz and 6\,Hz, 6\,Hz and 8\,Hz, 8\,Hz and 10\,Hz, \dots, until 38\,Hz and 40\,Hz. It has been proven that these frequency bands yield the best performance~\cite{Hersche2018FastFeatures,wang2021mrc}.
}

Afterwards, the spatial covariance matrix $\mathbf{C}_k \in \mathbb{R}^{N_{ch} \times N_{ch}}$ for each frequency band can be estimated as 
\begin{equation}\label{eq:riemann:design:covmat}
  \mathbf{C}_k = \hat{\mathbf{X}}_k \hat{\mathbf{X}}^T_k + \rho \mathbf{I},
\end{equation}
with $\rho$ being the regularization parameter that scales the identity matrix $\mathbf{I}$. 
\rebuttal{These covariance matrices are by construction symmetric and, with sufficient data, positive definite. A correct way to manipulate them for the final classifier is to rely on Riemannian geometry~\cite{BARACHANT2013whitening}. According to which, each covariance matrix can be projected locally onto an Euclidean tangent plane, 
which is done by the next operation: 
}
\begin{equation}\label{eq:riemann:design:whitening}
  \mathbf{W}_k = \mathbf{C}_{\text{ref},k}^{-1/2} \mathbf{C}_k \mathbf{C}_{\text{ref},k}^{-1/2}.
\end{equation}
\rebuttal{The covariance matrix is transformed by multiplying it with a reference matrix $\mathbf{C}_{\text{ref},k}^{-1/2}$, which is}
obtained beforehand during the model training by averaging the covariance matrices of all the training data for each frequency band.
\rebuttal{The mapping of the covariance matrix to the tangent space can be interpreted as a whitening operation~\cite{BARACHANT2013whitening}; hence, we name it \emph{Whitening} in this paper as a shorthand.}

The matrix logarithm is subsequently applied on the \emph{whitened} matrices $\mathbf{W}_k$ for finalizing the data mapping:
\begin{equation}\label{eq:riemann:design:logm}
  \mathbf{L}_k = \text{logm}(\mathbf{W}_k).
\end{equation}
The function $\text{vect}(\mathbf{L}_k)$ follows and vectorizes the output $\mathbf{L}_k$ of the matrix logarithm. $\mathbf{L}_k$ is a symmetric matrix and its upper right off-diagonal elements are multiplied by $\sqrt{2}$ to preserve the norm. The diagonal values $l_{k,(i,i)}$ and the upper-right off-diagonal elements $l_{k,(i,j)}$, with $i \in {1,2,...,N_{ch}}$ and $j \in {2,3,...,N_{ch}}$, are concatenated to form a vector:
\begin{equation}\label{eq:riemann:design:vect}
  \mathbf{v}_k = \text{vect}(\mathbf{L_k}) = [l_{k,(i,i)}] \oplus [\sqrt{2} \cdot l_{k,(i,j)}]. 
\end{equation}
The features vectors $\mathbf{v}_k$ from all frequency bands are finally concatenated and fed as input to the classifier. 

}
\new{
\subsection{Classification} \label{sec:model_classifier}
Based on the obtained feature vector, a simple classifier then decides whether the person was thinking about moving the `left hand', `right hand', `both feet', or `tongue', using the BCI Competition IV-2a dataset~\cite{Brunner2008BCIA}. We analyze two scenarios, a one-fits-all solution and a trade-off analysis of the average accuracy against the memory requirements. 

For the former, we train a linear \gls{svm} on the features extracted for each patient's training data and evaluate the accuracy of the overall pipeline with and without quantization (following the quantization described in Sec.~\ref{ch:riemann:implementation}). 

For the latter, we search for linear \glspl{svm} and \glspl{mlp} for each patient, tuning either the SVM's L2 regularization parameter or the MLP's number of hidden layers (between 1 and 3), the activation function (ReLU or sigmoid), and the number of nodes for each hidden layer separately in $\{2,4,8,\dots,2048\}$. 
}

\begin{figure}[t]
    \centering
    \includegraphics[width=\columnwidth]{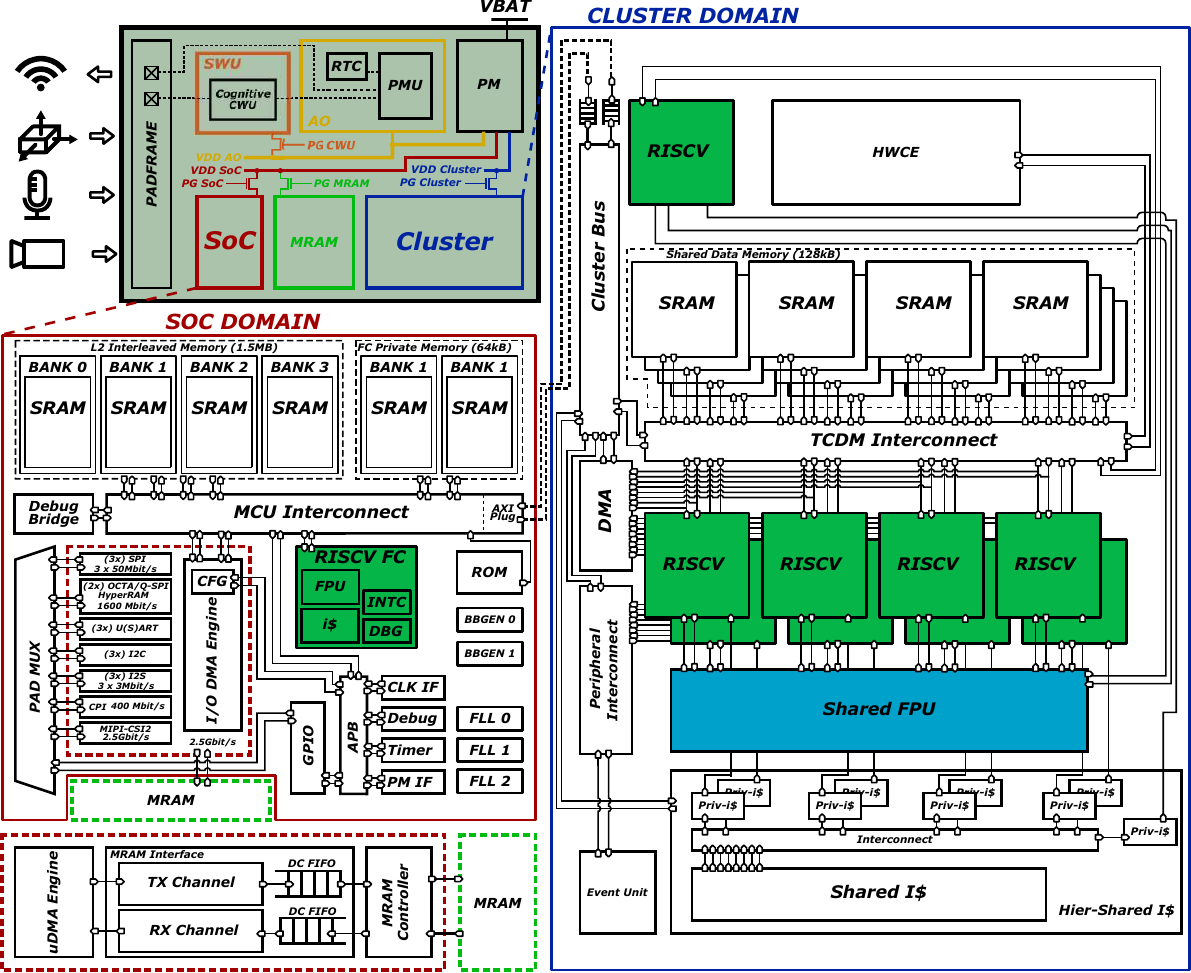}
    \caption{\gls{soc} architecture and power domains of \vega. Adapted from~\cite{Rossi2021vega}.}
    \label{fig:vega}
\end{figure}

\new{
\section{Parallel Ultra-Low Power Processor}

In this section we summarize the key features of the \gls{pulp} \gls{mcu} used in this work, called \vega~\cite{Rossi2021vega}.
It is designed for near-sensor applications featuring two main power domains, i.e., \gls{soc} and cluster, with 10 \mbox{RISC-V} cores, in addition to an always-on domain with a \textit{cognitive} wake-up unit, as depicted in Fig.~\ref{fig:vega}. It operates from \mbox{0.6\,V -- 0.8\,V} and can be scaled from a fully retentive \textit{cognitive} sleep mode consuming only 1.7\,\textmu W up to 32.2\,GOPS peak performance consuming 49.4\,mW.

\vega features one \mbox{RISC-V} core in the \gls{soc} domain, called \gls{fc}, and nine parallel cores in the cluster. The \gls{fc} handles the several peripherals of the \gls{mcu} and executes simple computations. For compute-intensive tasks, the cluster is activated and the tasks are offloaded to the nine parallel cores with four-pipeline stages. Four shared \glspl{fpu} are available in the cluster for FP32, FP16, and bfloat floating-point operations, while a multi-banked L1 memory of size 128\,kB is shared among the cores and connected through a 1-cycle \gls{tcdm} interconnect for fast access. A hierarchical program cache serves the cores with 512\,B private per-core plus 4\,kB of 2-cycle-latency shared cache to maximize efficiency.
A \gls{dma} unit, orchestrated by one of the nine cores, can autonomously transfer data between the L1 memory and the much larger L2 memory present in the \gls{soc}, which is a multi-banked \gls{sram} of size 1.5\,MB and 64\,kB private memory for the \gls{fc}.
An additional \gls{dma} engine in the \gls{soc} can independently transfer I/O data between the L2 in the \gls{soc} and the 4MB of non-volatile \gls{mram} and an eventual off-chip HyperRAM. Thanks to the \gls{dma} units, the computation and the data transfer can be fully overlapped.

The \mbox{RISC-V} cores are based on RVC32IMF \gls{isa} with additional custom extensions named Xpulp for near-sensor analytics algorithms. The extensions include hardware loops, post-incremental load and store, 2-way and 4-way \gls{simd} operations respectively for 16-bit and 8-bit data types. Together with the 9-core parallelism and optimized memory hierarchy, \vega can deliver up to 15.6 8-bit GOPS and up to 612\,GOPS/W, and up to 3.3\,GFLOPS and 129\,GFLOPS/W~\cite{Rossi2021vega}. 

}

\section{Quantization and Implementation} \label{ch:riemann:implementation}

\new{

Unlike the feed-forward \glspl{cnn}, which can be quantized to very low numerical precision, e.g., down to 8 bits or beyond without accuracy loss \cite{jacob2018quantization,DBLP:journals/corr/abs-2001-01091,DBLP:journals/corr/abs-1905-10452}, the quantization of the \gls{mrc} may cause significant accuracy degradation, mainly due to the \rebuttal{numerical stability of the \gls{iir} filters and the computation of the matrix logarithm.}
The accuracy has to be preserved, while at the same time lower precision yields less memory footprint and more energy-efficient implementation by exploiting the \mbox{2-} and \mbox{4-way} \gls{simd} instructions of the hardware architecture.
Hence, a mixed-precision quantization is proposed with a combination of \mbox{8-,} \mbox{16-,} and 32-bit fixed-point and 32-bit floating-point representations, as depicted in \figref{fig:riemann:feature_extraction_quant}.}
{Moreover, to fully exploit the multi-core nature of the \vega \gls{mcu}, we implement the \gls{mrc} by splitting the feature extraction into two separate blocks. As \figref{fig:background:riemannian} highlights, the features of every frequency band can be computed individually. We divide the computation between the 9 cores as follows:}
\begin{enumerate}
    \item {The first part of the feature extraction (in blue) computes an \gls{iir} filter, followed by several matrix operations. Each \gls{iir} filter cannot be parallelized, as its implementation keeps state while moving along the signal. Instead, we can perform the filtering of the $N_{ch}$ different EEG channels in parallel. The subsequent matrix operations can then be computed using their optimized, parallel implementations. Hence, we use all 9 cores to compute the whitened signal $W_k$ for a single frequency band $k$, and repeat this for all $n=18$ bands.}
    \item {The second part of the feature extraction (in red) requires computation of the \gls{evd}. Parallel implementations of such algorithms do not yield a speedup factor close to the theoretical maximum, Hence, we utilize the multi-core nature of the \gls{mcu} by computing multiple decompositions in parallel, each of them assigned to a single processing core.}
    \item {The final classification (in green) is computed on a single core since it accounts for a negligible part of the execution time.}
\end{enumerate}
\new{In the following paragraphs, we elaborate on the details of the quantization and the implementation on \vega.}

\new{
\subsection{Quantization Terminology}\label{sec:methods:terminology}
Assume a value $x \in \mathbb{R}$, the fixed-point representation $\tilde{x} \in \mathbb{Z}$ (in two's complement) of $x$ is written as
\begin{equation}\label{eq:background:quant:full}
  \tilde{x} = \text{clip}\left( \text{round}\left( x \cdot \frac{R_x}{S_x} \right), -R_x, R_x - 1 \right)
\end{equation}
with $R_x = 2^{n_x - 1}$,
or simplified, ignoring the clip and round function,
\begin{equation} \label{eq:background:quant:simple}
  \tilde{x} = x \cdot \frac{R_x}{S_x}.
\end{equation}
$S_x$ represents the \textit{dynamic range} of $x$ which can be represented by $\tilde{x}$.
If $\abs{x} > S_x$, then $\tilde{x}$ is clipped.
Choosing the dynamic range $S_x$ to be a power of two results in a traditional fixed-point format.
$R_x$ represents the number of quantization levels.
If $\tilde{x}$ is represented with $n_x$ bits, then $R_x = 2^{n_x - 1}$.
The fraction $R_x / S_x$ is called the \textit{conversion factor}.
}

\subsection{\gls{iir} Bandpass Filters}\label{ch:riemann:design:iir}
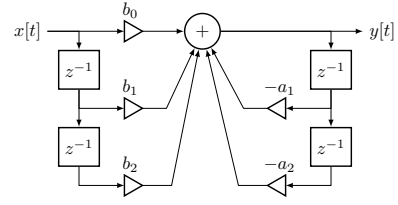
\begin{figure}
  \centering
    \centering
      \resizebox{0.6\columnwidth}{!}{%
    \begin{tikzpicture}[node distance=1.5cm, scale=1, transform shape]
      \node (input) {$x[t]$};
      \node[block, right of=input, yshift=-0.75cm, xshift=-0.5cm] (d11) {$z^{-1}$};
      \node[block, below of=d11] (d12) {$z^{-1}$};
      \node[gain, right of=input, xshift=0.5cm, label=above:{$b_0$}] (b0) {};
      \node[gain, below of=b0, label=above:{$b_1$}] (b1) {};
      \node[gain, below of=b1, label=above:{$b_2$}] (b2) {};
      \node[sum, right of=b0, xshift=0.4cm] (sum) {$+$};
      \node[revgain, right of=sum, yshift=-1.5cm, label=above:{$-a_1$}] (a1) {};
      \node[revgain, below of=a1, label=above:{$-a_2$}] (a2) {};
      \node[block, right of=a1, xshift=-0.5cm, yshift=0.75cm] (d21) {$z^{-1}$};
      \node[block, below of=d21] (d22) {$z^{-1}$};
      \node[right of=sum, xshift=2cm] (output) {$y[t]$};

      \draw[->] (input) -| (d11);
      \draw[->] (input) -- (b0);
      \draw[->] (d11) -- (d12);
      \draw[->] (d11) |- (b1);
      \draw[->] (d12) |- (b2);
      \draw[->] (b0) -- (sum);
      \draw[->] (b1) -- ++(0.7, 0) -- (sum);
      \draw[->] (b2) -- ++(0.8, 0) -- (sum);
      \draw[->] (sum) -- (output);
      \draw[->] (sum) -| (d21);
      \draw[->] (d21) -- (d22);
      \draw[->] (d21) |- (a1);
      \draw[->] (d22) |- (a2);
      \draw[->] (a1) -- ++(-0.7, 0) -- (sum);
      \draw[->] (a2) -- ++(-0.8, 0) -- (sum);
    \end{tikzpicture}
    }
    \caption{IIR filter realization: Direct Form I as defined in~\cite{smith2020filter}.}\label{fig:riemann:iir:df1}
\end{figure}

\new{
The input data $\mathbf{X}$ is quantized to 8 bits and is filtered with \gls{iir} filters.
Unlike \gls{fir} filters, i.e., convolutions, that are always numerically and asymptotically stable,} the \gls{iir} filters can become unstable, especially when they are quantized.
The internal accumulators can diverge, even if the output remains bounded.
\new{Different realization forms exist~\cite{smith2020filter}, yielding different asymptotic behavior of the internal signals and thus influencing the numerical stability.}
We implement the Direct-Form I, \new{shown in Fig.~\ref{fig:riemann:iir:df1}.}
It ensures that no internal overflows happen, because all internal registers store either the input or the output of the filter~\cite{smith2020filter}.
A typical approach for quantizing an \gls{iir} filter is to express them as a cascade of \glspl{sos}. 
\new{In this representation, multiple separate second-order \gls{iir} filters are cascaded allowing different sections to be quantized with different dynamic ranges}, thus minimizing the effect of quantizing the filter coefficients on the impulse response.
\rebuttal{We implement the 4th order Butterworth filters as a cascade of two \glspl{sos} with the following transfer function:
\begin{equation}
  H(z) = \prod_{m=0}^{M-1}\frac{b_{0,m} + b_{1,m} z^{-1} + b_{2,m} z^{-2}}{1 + a_{1,m} z^{-1} + a_{2,m} z^{-2}},
\end{equation}
where $z$ is a complex variable, $M$ is the number of sections, $a$ and $b$ are respectively the reverse and forward coefficients as depicted in Fig.~\ref{fig:riemann:iir:df1}. There are a total of $18 \cdot 2 \cdot 5 = 180$ coefficients, which we release together with the code\footref{footnote:code}.}

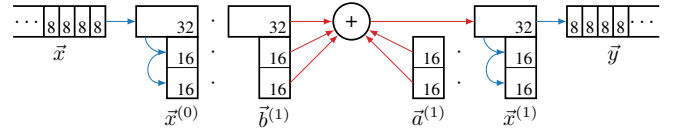
\begin{figure}
  \centering
    \resizebox{\columnwidth}{!}{%
  \begin{tikzpicture}[scale=1, transform shape]
    \draw[thick] (1, 1.5) -- (2.5, 1.5) -- (2.5, 1) -- (1, 1)
                 (1.5, 1.5) -- (1.5, 1)
                 (1.75, 1.5) -- (1.75, 1)
                 (2, 1.5) -- (2, 1)
                 (2.25, 1.5) -- (2.25, 1);
    \draw (1.75, 1) node[below] {$\vec{x}$};
    \draw (1.25, 1.25) node {$\dots$};
    \draw (2.125, 1.15) node {\footnotesize 8};
    \draw (2.375, 1.15) node {\footnotesize 8};
    \draw (1.625, 1.15) node {\footnotesize 8};
    \draw (1.875, 1.15) node {\footnotesize 8};

    \draw[thick] (3,1) rectangle (4, 1.5);
    \draw[thick] (3.5,0) rectangle (4, 1);
    \draw (3.5, 0.5) -- (4, 0.5);
    \draw (3.8, 1.15) node{\footnotesize 32};
    \draw (3.8, 0.15) node {\footnotesize 16};
    \draw (3.8, 0.65) node {\footnotesize 16};
    \draw (3.75, 0) node[below]{$\vec{x}^{(0)}$};

    \draw (4.25, 0.75) node {$\cdot$};
    \draw (4.25, 0.25) node {$\cdot$};
    \draw (4.25, 1.25) node {$\cdot$};
    
    \draw[thick] (4.5,1) rectangle (5.5, 1.5);
    \draw[thick] (5,0) rectangle (5.5, 1);
    \draw (5, 0.5) -- (5.5, 0.5);
    \draw (5.3, 1.15) node{\footnotesize 32};
    \draw (5.3, 0.15) node {\footnotesize 16};
    \draw (5.3, 0.65) node {\footnotesize 16};
    \draw (5.25, 0) node[below]{$\vec{b}^{(1)}$};

    \node[sum] (acc1) at (6.5, 1.25) {+};

    \draw[thick] (7.5,0) rectangle (8, 1);
    \draw (7.5, 0.5) -- (8, 0.5);
    \draw (7.8, 0.15) node {\footnotesize 16};
    \draw (7.8, 0.65) node {\footnotesize 16};
    \draw (7.76, 0) node[below]{$\vec{a}^{(1)}$};

    \draw (8.25, 0.75) node {$\cdot$};
    \draw (8.25, 0.25) node {$\cdot$};

    \draw[thick] (8.5,1) rectangle (9.5, 1.5);
    \draw[thick] (9,0) rectangle (9.5, 1);
    \draw (9, 0.5) -- (9.5, 0.5);
    \draw (9.3, 1.15) node{\footnotesize 32};
    \draw (9.3, 0.15) node {\footnotesize 16};
    \draw (9.3, 0.65) node {\footnotesize 16};
    \draw (9.25, 0) node[below]{$\vec{x}^{(1)}$};

    \draw[thick] (11.5, 1) -- (10, 1) -- (10, 1.5) -- (11.5, 1.5)
                 (10.75, 1.5) -- (10.75, 1)
                 (11, 1.5) -- (11, 1)
                 (10.25, 1.5) -- (10.25, 1)
                 (10.5, 1.5) -- (10.5, 1);
    \draw (10.75, 1) node[below] {$\vec{y}$};
    \draw (11.25, 1.25) node {$\dots$};
    \draw (10.625, 1.15) node {\footnotesize 8};
    \draw (10.875, 1.15) node {\footnotesize 8};
    \draw (10.125, 1.15) node {\footnotesize 8};
    \draw (10.375, 1.15) node {\footnotesize 8};
    
    \draw[->, color0] (2.5, 1.25) -- (3, 1.25);
    \draw[->, color0] (3.15, 1) to[out=-90, in=180] (3.5, 0.75);
    \draw[->, color0] (3.5, 0.75) to[out=180, in=180, looseness=2] (3.5, 0.25);
    
    \draw[->, color0] (8.65, 1) to[out=-90, in=180] (9, 0.75);
    \draw[->, color0] (9, 0.75) to[out=180, in=180, looseness=2] (9, 0.25);
    \draw[->, color0] (9.5, 1.25) -- (10, 1.25);

    \draw[->, color3] (5.5, 1.25) -- (acc1);
    \draw[->, color3] (5.5, 0.75) -- (acc1);
    \draw[->, color3] (5.5, 0.25) -- (acc1);
    \draw[->, color3] (7.5, 0.75) -- (acc1);
    \draw[->, color3] (7.5, 0.25) -- (acc1);
    \draw[->, color3] (acc1) -- (8.5, 1.25);

  \end{tikzpicture}
  }
  \caption{IIR filter realization and implementation. Diagram showing the registers and the operations of a single \gls{sos}, where arrows in blue represent register transfers and shuffle operations, and red arrows represent \gls{mac} instructions. For an \gls{iir} filter with two \gls{sos} the output $\vec{x}^{(1)}$ can be reused. 
  }\label{fig:riemann:iir_reg}
\end{figure}

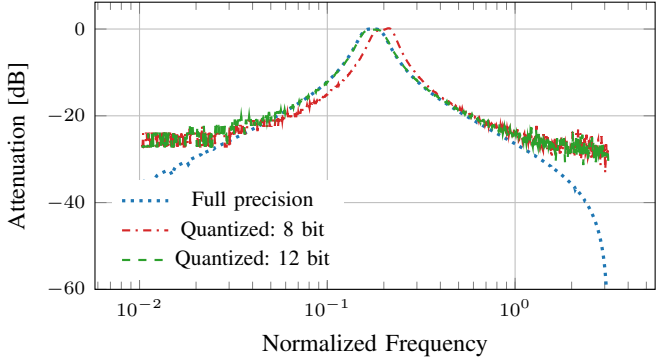
\begin{figure}[t]
  \centering
  \resizebox{\columnwidth}{!}{%
  \begin{tikzpicture}
    \begin{axis}[
      no markers,
      xmode=log,
      grid,
      width=\columnwidth,
      height=0.6\columnwidth,
      legend pos=south west,
      xlabel={\small Normalized Frequency},
      ylabel={\small Attenuation [dB]},
      ymin = -60,
      legend style={
        draw=none,
        font={\footnotesize}
      },
      ]
      \addplot+ [very thick, dotted] table [x=f, y=float, col sep=comma] {measurements/iir_sweep_0.048-0.064_N8.csv}; 
      \addplot+ [thick, dashdotted] table [x=f, y=quant, col sep=comma, ] {measurements/iir_sweep_0.048-0.064_N8.csv};
      \addplot+ [thick, dashed] table [x=f, y=quant, col sep=comma] {measurements/iir_sweep_0.048-0.064_N12.csv};
      \legend{Full precision, Quantized: 8 bit, Quantized: 12 bit};
    \end{axis}
  \end{tikzpicture}
  }
  \vspace{-0.5cm}
  \caption{Frequency response of \rebuttal{a 4th order Butterworth \gls{iir} band-pass filter with lower and upper cut-off frequencies of 6\,Hz and 8\,Hz, respectively.} The full precision filter (blue dotted line) is compared with the quantized implementations, where the filter coefficients are quantized to 8 (red dash-dotted line) and 12 (green dashed line) bits, and the output to 8 bits.
  }\label{fig:riemann:filter:frequency}
\end{figure}

\begin{figure}[t]
  \centering
  \resizebox{\columnwidth}{!}{%
  \begin{tikzpicture}
    \begin{axis}[
      no markers,
      grid,
      width=\columnwidth,
      height=0.6\columnwidth,
      legend pos=south west,
      xlabel={\small Frequency [Hz]},
      ylabel={\small Attenuation [dB]},
      xmin = 2.5,
      xmax = 13.5,
      ymin = -12,
      legend style={
        draw=none,
        font={\footnotesize}
      },
      xtick = {4,6,8,10,12},
      xticklabels = {4Hz, 6Hz, 8Hz, 10Hz, 12Hz},
      ]
      
      \addplot+ [black,      thick, dashed] coordinates {(0, 0)};
      \addplot+ [black, very thick, dotted] coordinates {(0, 0)};
      \addplot+ [black,      thick]         coordinates {(0, 0)};
      \legend{Full precision, Quantized: 8 bit, Quantized: 12 bit};
      
      \draw[draw=none, fill=matplotlib0, opacity=0.1] (axis cs: 4,5) rectangle (axis cs: 6,-20);
      \draw[draw=none, fill=matplotlib1, opacity=0.1] (axis cs: 6,5) rectangle (axis cs: 8,-20);
      \draw[draw=none, fill=matplotlib2, opacity=0.1] (axis cs: 8,5) rectangle (axis cs:10,-20);
      \draw[draw=none, fill=matplotlib3, opacity=0.1] (axis cs:10,5) rectangle (axis cs:12,-20);
      
      \addplot+ [matplotlib0,      thick, dashed] table [x=f, y={0.032-0.048_float}, col sep=comma] {measurements/iir_sweep_all_log.csv};
      \addplot+ [matplotlib0, very thick, dotted] table [x=f, y={0.032-0.048_q8},    col sep=comma] {measurements/iir_sweep_all_log.csv};
      \addplot+ [matplotlib0,      thick]         table [x=f, y={0.032-0.048_q12},   col sep=comma] {measurements/iir_sweep_all_log.csv};
      \addplot+ [matplotlib1,      thick, dashed] table [x=f, y={0.048-0.064_float}, col sep=comma] {measurements/iir_sweep_all_log.csv};
      \addplot+ [matplotlib1, very thick, dotted] table [x=f, y={0.048-0.064_q8},    col sep=comma] {measurements/iir_sweep_all_log.csv};
      \addplot+ [matplotlib1,      thick]         table [x=f, y={0.048-0.064_q12},   col sep=comma] {measurements/iir_sweep_all_log.csv};
      \addplot+ [matplotlib2,      thick, dashed] table [x=f, y={0.064-0.080_float}, col sep=comma] {measurements/iir_sweep_all_log.csv};
      \addplot+ [matplotlib2, very thick, dotted] table [x=f, y={0.064-0.080_q8},    col sep=comma] {measurements/iir_sweep_all_log.csv};
      \addplot+ [matplotlib2,      thick]         table [x=f, y={0.064-0.080_q12},   col sep=comma] {measurements/iir_sweep_all_log.csv};
      \addplot+ [matplotlib3,      thick, dashed] table [x=f, y={0.080-0.096_float}, col sep=comma] {measurements/iir_sweep_all_log.csv};
      \addplot+ [matplotlib3, very thick, dotted] table [x=f, y={0.080-0.096_q8},    col sep=comma] {measurements/iir_sweep_all_log.csv};
      \addplot+ [matplotlib3,      thick]         table [x=f, y={0.080-0.096_q12},   col sep=comma] {measurements/iir_sweep_all_log.csv};
      
    \end{axis}
  \end{tikzpicture}
  }
  \vspace{-0.5cm}
  \caption{\rebuttal{A zoomed-in illustration of the frequency responses. The quantization to 8 bits causes overshoot (blue), undershoot (orange), or shift (red and green) of the frequency bands. The quantization to 12 bits has minimal effect compared to the full precision filters. Four out of 18 filters are illustrated, while the others present similar behavior. } }\label{fig:riemann:filter:frequency_all}
\end{figure}
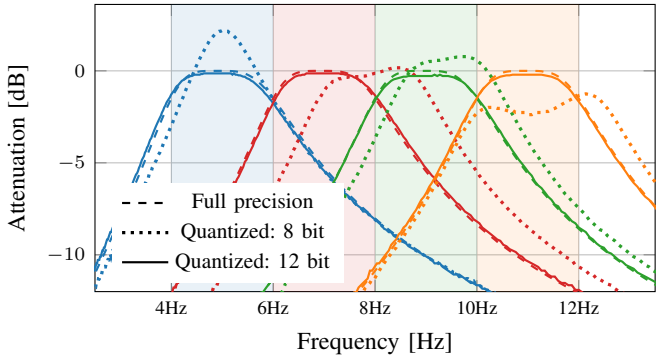

\rebuttal{
The choice of the quantization levels depends on three requirements: a) The quantization must not change the frequency response of the filters; b) No numerical overflow must happen; c) The underlying hardware can be optimally used.}
\new{
\rebuttal{
Quantizing all filter coefficients with 8 bits has a significant impact on the impulse response of the filter, as shown in Fig.~\ref{fig:riemann:filter:frequency} and Fig.~\ref{fig:riemann:filter:frequency_all}, where we observe overshooting, undershooting, or shifting behaviours of the frequency responses compared to the full-precision ones. We increase the number of bits and observe that a 12-bit representation has unnoticeable effect on the frequency responses which closely overlap with the full-precision ones. 
Another strictly related parameter is the number of bits used for the internal signals. We observe that using 12 bits for the filter coefficients and 16-bit registers to accumulate intermediate values \rebuttal{does} not cause any overflow for the dataset of interest, and it allows the usage of 2-way \gls{simd} operations for the following iteration of the filter computation yielding a more efficient implementation. Thus, we choose 12 bits to quantize the filter coefficients, since a higher number of bits is not necessary for preserving the frequency responses and use 16 bits to represent the intermediate results of the \gls{sos}, as depicted in Fig.~\ref{fig:riemann:iir_reg}.
Each \gls{sos} contains three \glspl{mac} for the forward accumulation 
and two \glspl{mac} for the backward accumulation. 
This enables the usage of \gls{simd} instructions with bit-width 16.
}

Moreover, the dynamic ranges of the coefficients of each \gls{sos} are chosen independently to minimize the quantization error and set to be a power of two, such that no expensive divisions are required with the advantage of using more energy-efficient bit-shift operations. 
}
\new{The \riscy{} cores on \vega support instructions (namely \lstinline!p.addNR! and \lstinline!p.subNR!), which normalize and round the result of an addition or a subtraction accordingly. 
Therefore, all intermediate results can be rounded efficiently, improving the precision of the filter.
For the parallel computation, we assign each 1D signal of size $N_s$ from the $N_{ch}$ \gls{eeg} electrodes to individual cores for the filtering. In other words, $\hat{\mathbf{X}}_k$ is computed in parallel using the nine cores. Once finished, $\hat{\mathbf{X}}_{k+1}$ is subsequently assigned to the cluster for the parallel computation.
}

\subsection{Covariance Matrix} \label{ch:riemann:implementation:design:covmat}

\new{
Following the terminology introduced in Sec.~\ref{sec:methods:terminology}, the covariance matrix $\mathbf{C}_k$ is quantized as
\begin{equation}
  \tilde{\mathbf{C}}_k = \frac{\tilde{\mathbf{X}}_k^T \tilde{\mathbf{X}}_k + \tilde{\rho} \mathbf{I}}{\frac{R_x^2 S_c}{S_x^2 R_c}} = \frac{\tilde{\mathbf{X}}_k^T \tilde{\mathbf{X}}_k + \tilde{\rho} \mathbf{I}}{F_c},
\end{equation}
with $F_c = \frac{2^{2n_x - 2} \cdot S_c}{2^{n_c - 1} \cdot S_x^2}$, $\tilde{\mathbf{X}}_k$ being the filtered signals represented in 8 bits, $R$ the quantization levels, and $S$ the dynamic range. We force $S_c / S_x^2$ to be a power of two, such that the scaling factor $F_c$ is also a power of two, allowing us to implement the transformation with a bit-shift. Note that $\tilde{\mathbf{C}}_k$ is a symmetric matrix, hence} we compute only the upper-right triangle and copy the remaining elements. 4-way \gls{simd} operations are used for the 8-bit numbers and the computation is implemented concurrently by splitting the upper-right part of the output matrix among all processing units.

\subsection{Whitening}\label{sec:methods:whitening}
\new{
We notice that the input to the matrix logarithm is sensitive to quantization errors \rebuttal{due to the calculation of the \gls{evd}~\cite{Wilkinson1962EVDerror}}. Thus, to avoid accuracy degradation, we use the high dynamic range available with 32 bits for $\mathbf{W}_k$.
Let $n_c$ and $n_{\text{ref}}$ be the number of bits to represent $\tilde{\mathbf{C}}_k$ and $\tilde{\mathbf{C}}_{\text{ref},k}^{-1/2}$, respectively. We rescale $\tilde{\mathbf{C}}_k$ to use the entire dynamic range with $n_c$ bits, while the reference matrix $\mathbf{C}_{\text{ref},k}^{-1/2}$ is quantized to $n_{\text{ref}}$ bits during training. We do not scale either the intermediate, nor the final result of Eq.~\ref{eq:riemann:design:whitening} to minimize the quantization error.}
Since we can exploit either \mbox{4-} or 2-way \gls{simd} operations, we test both $n_c = n_{\text{ref}} = 8$ and $16$. The accuracy drops significantly with the former, while the latter causes overflows. Hence, we reduce $n_{\text{ref}}$, until training completes without overflow, resulting in $n_{\text{ref}} = 11$.
With $n_c = 16$ and $n_{\text{ref}} = 11$ we obtain similar accuracy to the full-precision version.
\new{The dynamic range $S_w$ is determined by
\begin{equation}
  \tilde{\mathbf{W}} \frac{R_w}{S_w} = \tilde{\mathbf{C}}_{\text{ref}}^{-1/2} \tilde{\mathbf{C}} \tilde{\mathbf{C}}_{\text{ref}}^{-1/2}\frac{R_c R_{\text{ref}}^2}{S_c S_{\text{ref}}^2},
\end{equation}
which means
\begin{equation}
  S_w = \frac{2^{n_w - 1} \cdot S_c S_{\text{ref}}^2}{2^{n_c - 1} \cdot 2^{2n_{\text{ref}}-2}}.
\end{equation}
The result of the Whitening block is then multiplied by $S_w / 2^{n_w-1}$ to be converted to floating-point representation as input to the matrix logarithm.

We implement the two \glspl{mmm} of the Whitening operation using the PULP-DSP library~\cite{wang2019dsp}. The first multiplication is computed in 16 bits by exploiting the \gls{simd} instructions, the second in 32 bits to preserve the full dynamic range. {The DSP library splits the computation into small chunks of size $2 \times 4$ in the 16-bit implementation, and $2 \times 2$ in the 32-bit version. Each chunk is assigned to a single core to be computed.}
}

\subsection{Matrix Logarithm}\label{ch:riemann:design:logm}

\new{
\begin{algorithm}[b]
  \caption{Householder Tridiagonalization~\cite{burden2004numerical}}\label{alg:householder}
  \begin{algorithmic}[1]\small
    \Procedure{HhTridiag}{$\mathbf{A}$}\Comment{$\mathbf{A} \in \mathbb{R}^{n \times n}$ with elements $a_{j,k}$}
      \State{$\mathbf{Q} \leftarrow \mathbf{I}$}
      \For{$k \in \{ 1, 2, \ldots, n-2 \}$}\Comment{working on column $k$}
        \State{$\alpha \leftarrow -\text{sgn}\left( a_{k+1,k} \right) \cdot \norm{\mathbf{A}_{k+1:n,k}}$}\Comment{euclidean norm}
        \State{$r \leftarrow \sqrt{\frac{1}{2} \alpha \left( \alpha - a_{k+1,k} \right)}$}
        \State{$\vec{v} \leftarrow {\left[0, \dots, 0, a_{k+1,k} - \alpha, a_{k+2,k}, \dots, a_{n,k}\right]}^T / 2r$}
        \State{$\mathbf{P} \leftarrow \mathbf{I} - 2 \vec{v} \vec{v}^T$}\Comment{Householder matrix}
        \State{$\mathbf{A} \leftarrow \mathbf{P} \mathbf{A} \mathbf{P}^T$}\label{alg:householder:a_step}
        \State{$\mathbf{Q} \leftarrow \mathbf{Q} \mathbf{P}$}\label{alg:householder:q_step}
      \EndFor
      \State{\textbf{return} $\mathbf{A},\ \mathbf{Q}$}\Comment{$\mathbf{A}$ tridiagonal, $\mathbf{Q}$ orthogonal}
    \EndProcedure
  \end{algorithmic}
\end{algorithm}
}

The matrix logarithm of a square matrix $\mathbf{A} \in \mathbb{R}^{n \times n}$ is defined in terms of its \gls{evd}, as
\begin{equation}
  \text{logm}(\mathbf{A}) = \mathbf{Q}^{-1} \text{logm}(\mathbf{D}) \mathbf{Q},
\end{equation}
where $\mathbf{A} = \mathbf{Q}^{-1} \mathbf{D} \mathbf{Q}$, \rebuttal{with $\mathbf{D}$ containing the Eigenvalues of $\mathbf{A}$ on its diagonal and $\mathbf{Q}$ containing the Eigenvectors of $\mathbf{A}$. The logarithm of a diagonal matrix $\mathbf{D}$ is computed by applying the logarithm to its diagonal element, i.e., 
\begin{equation}
\small
\arraycolsep1pt
  \text{logm}\left(
    \begin{bmatrix}
      d_1 & 0 & \dots & 0 \\
      0 & d_2 & \dots & 0 \\
      \vdots & \vdots & \ddots & \vdots \\
      0 & 0 & 0 & d_n \\
    \end{bmatrix}
  \right) = \begin{bmatrix}
    \log(d_1) & 0 & \dots & 0 \\
    0 & \log(d_2) & \dots & 0 \\
    \vdots & \vdots & \ddots & \vdots \\
    0 & 0 & 0 & \log(d_n) \\
  \end{bmatrix}.
\end{equation}
The matrix logarithm exists in $\mathbb{R}^{n \times n}$ only if $\mathbf{A} \succ 0$ is positive definite, because a positive definite matrix has only positive Eigenvalues which ensure the existence of logarithm in $\mathbb{R}$.
}

\new{
The feature extraction of the \gls{mrc} only computes the matrix logarithm of dense and symmetric matrices, meaning that we can optimize the \gls{evd} by computing first the tridiagonal decomposition which produces a tridiagonal matrix similar to the original one, i.e., the Eigenvalues are preserved. Computing the \gls{evd} of a tridiagonal matrix yields less computational effort. }
The final transformation is of the form
\begin{equation}
  \mathbf{A} = \mathbf{Q}_t^T \mathbf{T} \mathbf{Q}_t = \mathbf{Q}_t^T \mathbf{Q}_d^T \mathbf{D} \mathbf{Q}_d \mathbf{Q}_t,
\end{equation}
where $\mathbf{Q}_t$ is the orthogonal matrix for the tridiagonal transformation and $\mathbf{Q}_d$ the one for the \gls{evd}.
$\mathbf{Q}_d \mathbf{Q}_t$ is an orthogonal matrix containing the Eigenvectors of $\mathbf{A}$, 
\new{and $\mathbf{T}$ is of the form
\begin{equation}\label{eq:background:tridiagonal}\small
  \mathbf{T} = \begin{bmatrix}
    a_1 & b_2 & 0 & 0& \dots & 0 & 0 \\
    b_2 & a_2 & b_3 & 0& \dots & 0 & 0 \\
    0 & b_3 & a_3 & b_4& \dots & 0 & 0 \\
    0 & 0 & b_4 & a_4& \dots & 0 & 0 \\
    \vdots & \vdots & \vdots & \vdots & \ddots & \vdots  & \vdots\\
    0 & 0 & 0 & 0 & \dots & a_{n-1} & b_{n} \\
    0 & 0 & 0 & 0 & \dots & b_n & a_n \\
  \end{bmatrix}
\end{equation}
containing the same Eigenvalues of $\mathbf{A}$.}

To compute the tridiagonal matrix, we use the Householder transformation~\cite{burden2004numerical}, \new{reported in Algorithm~\ref{alg:householder}. 
This algorithm has complexity $\mathcal{O}\left(n^4\right)$, since it contains \glspl{mmm} (on line~\ref{alg:householder:a_step} and line~\ref{alg:householder:q_step}), which take $\mathcal{O}\left(n^3\right)$, for a total of $\mathcal{O}(n)$ iterations.
We can improve Algorithm~\ref{alg:householder} by grouping the operations differently:
\begin{align}
  \mathbf{A}^+ = \mathbf{P} \mathbf{A} \mathbf{P}^T &= \left(\mathbf{I} - 2 \vec{v} \vec{v}^T\right) \mathbf{A} \left( \mathbf{I} - 2 \vec{v} \vec{v}^T \right)\\
&= \mathbf{A} - 2\vec{v} \vec{v}^T \mathbf{A} - 2 \mathbf{A} \vec{v} \vec{v}^T + 4 \vec{v} \vec{v}^T \mathbf{A} \vec{v} \vec{v}^T
\end{align}
Recall that $\mathbf{A}$ is symmetric.
We can simplify this equation introducing the row vector $\vec{b}^T = \vec{v}^T \mathbf{A} \in \mathbb{R}^n$, the matrix $\mathbf{B} = \vec{v} \vec{b}^T \in \mathbb{R}^{n \times n}$ and the constant $c = \vec{b}^T \vec{v} \in \mathbb{R}$, and get:
\begin{equation}
  \mathbf{P} \mathbf{A} \mathbf{P}^T = \mathbf{A} - 2 (\mathbf{B} + \mathbf{B}^T) + 4c \vec{v} \vec{v}^T,
\end{equation}
with $\vec{b}^T = \vec{v}^T \mathbf{A},\ \mathbf{B} = \vec{v} \vec{b}^T,\ c = \vec{b}^T \vec{v}$.
This shows, that we can reduce the two matrix multiplications ($\mathcal{O}(n^3)$) to one vector matrix multiplication, one inner vector product, two outer vector products, two scalar matrix multiplications, and three matrix additions, resulting in a total complexity of $\mathcal{O}(n^2)$.
Similarly, we can compute the update for matrix $\mathbf{Q}$ to reduce the matrix-matrix multiplication into simpler operations of complexity $\mathcal{O}(n^2)$:
\begin{align}
  \mathbf{Q}^+ = \mathbf{Q} \mathbf{P} &= \mathbf{Q} (\mathbf{I} - 2 \vec{v} \vec{v}^T) \\
                                      &= \mathbf{Q} - 2 (\mathbf{Q} \vec{v}) \vec{v}^T.
\end{align}

Finally, we can use the fact that the vector $\vec{v}$ contains zeros up to the current iteration $k$ (see Algorithm~\ref{alg:householder}).
We use this to further reduce the computational complexity.
With $k' = k + 1$, in total, this new method has the complexity
\begin{equation}\label{eq:riemann:design:householder:complexity}
  \mathcal{O}\left(\sum_{k'=2}^{n-1} k' + 4k' n + 2k'^2\right) = \mathcal{O}\left( \frac{8 n^3}{3} \right).
\end{equation}
}

For computing the diagonal matrix $\mathbf{D}$ from the tridiagonal symmetric matrix $\mathbf{T}$, we use the QR algorithm with implicit Wilkinson Shift~\cite{wilkinson1965evd}. 
\new{This algorithm is an iterative process of applying a single QR step to the matrix $\mathbf{T}$, with the goal to bring each element on the off-diagonal close to zero.
The algorithm is designed such that the off-diagonal elements on the lower right are reduced first.
Once this element is smaller than $\epsilon$, the algorithm no longer considers this row and column of the matrix, effectively reducing the size of the matrix for the remaining iterations.
In some cases, the algorithm does not converge.
Assuming that the algorithm is currently working on off-diagonal element $b_m$, based on the notation of Eq.~\ref{eq:background:tridiagonal}, if an element $b_k$, with $k < m$, becomes zero before $b_m$ does, the QR update can no longer propagate through the entire matrix.
We fix this issue by splitting the matrix into two parts, namely $\mathbf{T}_{1:k-1,1:k-1}$ and $\mathbf{T}_{k:m,k:m}$, and solving the \gls{evd} separately in those regions by recursion.
The QR algorithm with implicit Wilkinson Shift has a complexity of $\mathcal{O}(6n^3)$~\cite{wilkinson1965evd}.
Combining this with the complexity of the Householder transformation derived in Eq.~\ref{eq:riemann:design:householder:complexity} yields a total complexity of $\mathcal{O}(9n^3)$.}

\new{
\rebuttal{Full-precision floating-point representation is necessary for computing the \gls{evd} for the algorithms to converge and to ensure numerical stability~\cite{Wilkinson1962EVDerror,wilkinson1965evd}.} Hence, we compute the matrix logarithm with floating-point values and exploit the \glspl{fpu} of \vega.} \rebuttal{In full-precision \gls{mrc}, the input of the matrix logarithm is always positive definite, ensuring the \rebuttal{existence} of the matrix logarithm. However, the Eigenvalues vary with the quantization of the input and in some cases even become negative, making it impossible to compute real logarithm~\cite{Wilkinson1962EVDerror}.} 
We address this issue by (a) making use of the entire 32-bit dynamic range for the inputs, as explained in Sec.~\ref{sec:methods:whitening}, and (b) clipping all Eigenvalues $\lambda_k$ to $\max\{\lambda_k, \lambda_{\min}\}$ by introducing a threshold $\lambda_{\min}=10^{-3}$ to ensure all Eigenvalues remain above zero. Its value is chosen based on the smallest Eigenvalue occurring while training the full precision \gls{mrc}.

For computing the \gls{evd}, we implement both the basic version of Householder transformation and the improved version~\cite{burden2004numerical} for speedup analyses.
The computation of the rotation matrix required for the Givens rotation~\cite{givens1954numerical} of each QR step is done exclusively with multiplications, divisions, and additions, without using expensive trigonometric functions~\cite{bindel2002computing}.
{For parallel implementation, every core is assigned with a frequency band and computes the Householder transformation and QR algorithm, i.e., nine matrix logarithms are computed concurrently by the nine cores of \vega. Once a core has finished with a frequency band, a next workload is assigned to it until all 18 matrix logarithms are computed. With nine cores, each core computes two matrix logarithms. }
Finally, we convert the results back to 8-bit fixed-point format using the dynamic range learned during training. 

\new{
\subsection{Classifier}
The final classifier in \gls{mrc} is a \gls{svm} or MLP, which we train on the quantized features.
The weights and biases are then quantized with bit-widths $n_w = 8$ and $n_b = 32$, respectively, by determining the dynamic ranges after training. We do not rescale the output because the prediction is based on the relative largest output value. Hence, the weight vectors can use the entire range available with 8 bit, reducing the quantization error.
The matrix-vector product of the \gls{svm} is computed using 8-bit \gls{simd} instructions.
We implement it on a single core, since it accounts for a negligible portion of the computation of the entire model.
In case of the MLP, parallel fast inference and quantization to 8-bit representation are discussed in-depth in \cite{wang2019fann,garofalo2020pulp}. 
}

\new{

\subsection{Power Measurements}

We measure the power consumption of \vega using the Keysight N6705C power analyzer with a sampling rate of 0.06144\,ms. Both \gls{soc} and cluster domains are measured in addition to the always-on domain. For comparison with~\cite{Schneider2020,wang2021mrc}, we measure with both the \gls{fc} and the cluster cores running at 50\,MHz, 100\,MHz, and 350\,MHz, supplied with the lowest (0.6\,V) and the highest (0.8\,V) voltages. Additionally, we vary the core frequencies for the two extreme supply voltages to find respectively the most energy-efficient and the fastest executions of the \gls{mrc} inference. {Finally, we fine-tune separately the frequencies of the \gls{soc} and the cluster domains to reach the optimal operating point.}

}

\section{Experimental Results and Discussion}\label{ch:results}

\begin{table*}[t]
  \centering
\new{
  \caption{\rebuttal{Classification accuracy (\%) on 2- (marked with $^{**}$) and 4-class \gls{mi} tasks for embedded implementation.}}\label{tab:results:accuracy}
    \small
    \begin{tabular}{lccccccccccc}
      \toprule
      & WOLA-CSP & \multicolumn{2}{c}{CSP+SVM} & \multicolumn{2}{c}{Q-\eegnet{}} & \multicolumn{6}{c}{\gls{mrc}} \\
      \cmidrule(r){1-1}
      \cmidrule(r){2-2}
      \cmidrule(r){3-4}
      \cmidrule(r){5-6}
      \cmidrule(){7-12}
      Ref. & \cite{bewalfi2018_wolacsp}$^{**}$ & \cite{malekmohammadi2019_cspsvm}$^{**}$ & \cite{malekmohammadi2019_cspsvm} & \cite{Schneider2020} & \cite{Schneider2020} & \cite{Hersche2018FastFeatures} & \cite{Hersche2018FastFeatures} &  ours & ours & ours & ours \\
      Classifier & LDA & \gls{svm} & \gls{svm} & built-in & built-in & lin. SVM & lin. SVM & lin. SVM & lin. SVM & variable & variable \\
      Precision & 16-bit & 20-bit & 20-bit & full & 8-bit & full & full & full & mixed & full & mixed \\
      \#features & -- & -- & -- & 200\,k & 200\,k & 32\,637 & 10\,879 & 4\,554 & 4\,554 & 4\,554 & 4\,554 \\
      \#temp. win. $t$ & -- & -- & -- & -- & -- & 3 & 1 & 1 & 1 & 1 & 1 \\
      \#freq. bands $f$ & -- & -- & -- & -- & -- & 43 & 43 & 18 & 18 & 18 & 18 \\
      Cov. Mat. Reg. $\rho$ & -- & -- & -- & -- & -- & 0.0 & 0.0 & 1.0 & 1.0 & 1.0 & 1.0 \\
      \cmidrule(r){1-1}
      \cmidrule(r){2-2}
      \cmidrule(r){3-4}
      \cmidrule(r){5-6}
      \cmidrule(){7-12}
      Subject 1 & 86.8 & 91.7 & 79.2 & 81.0 & 81.0 & 90.0 & 91.8 & 91.8 & 90.7 & svm: 91.8 & svm: 90.8\\
      Subject 2 & 63.9 & 59.7 & 56.3 & 57.6 & 53.1 & 55.5 & 51.6 & 53.7 & 51.2 & mlp: 60.4 & mlp: 57.6\\
      Subject 3 & 94.4 & 95.8 & 87.5 & 87.9 & 91.2 & 81.3 & 83.5 & 83.5 & 81.0 & svm: 83.5 & svm: 81.0\\
      Subject 4 & 68.8 & 77.1 & 63.9 & 61.6 & 58.1 & 71.9 & 73.3 & 73.7 & 74.1 & mlp: 74.6 & mlp: 74.1\\
      Subject 5 & 56.3 & 67.4 & 41.0 & 70.6 & 68.4 & 69.6 & 63.4 & 68.8 & 63.0 & svm: 68.8 & mlp: 63.8\\
      Subject 6 & 69.4 & 69.4 & 46.9 & 53.4 & 50.1 & 56.7 & 58.6 & 56.7 & 56.3 & mlp: 57.2 & svm: 56.3\\
      Subject 7 & 78.5 & 78.5 & 81.9 & 75.7 & 75.2 & 85.6 & 86.7 & 84.1 & 85.9 & svm: 87.0 & svm: 85.9\\
      Subject 8 & 97.9 & 97.2 & 76.0 & 77.4 & 81.2 & 83.8 & 81.6 & 81.5 & 82.7 & mlp: 82.3 & mlp: 83.4\\
      Subject 9 & 93.8 & 88.2 & 72.2 & 76.7 & 79.7 & 84.9 & 82.6 & 82.2 & 81.8 & mlp: 82.2 & svm: 81.8\\
      \cmidrule(r){1-1}
      \cmidrule(r){2-2}
      \cmidrule(r){3-4}
      \cmidrule(r){5-6}
      \cmidrule(){7-12}
      Avg. acc. & 78.85 & 80.55 & 67.21 & 71.3 & 70.9 & 75.5 & 74.8 &  \textbf{75.1} & \textbf{74.1} & \textbf{76.4} & \textbf{75.0}\\
      Std. dev. & -- & -- & -- & 11.5 & 14.3 & 12.8 & 13.9 & 12.2 & 13.2 & 11.3 & 12.0\\
      \bottomrule
    \end{tabular}
    }
\end{table*}

\setlength{\tabcolsep}{6.8pt} 
\begin{table*}[t]
  \centering
  \new{
  \caption{Compute time for \gls{mrc} on \vega. }
  \label{tab:results:riemann}
  {\small
    \begin{threeparttable}
    \begin{tabular}{@{}lrrrrrrcr@{}}
      \toprule
      & \multicolumn{2}{c}{baseline} & \multicolumn{2}{c}{improved \gls{evd}} & \multicolumn{2}{c}{parallel execution} & parallel & ops/cycle$^\nmid$ \\
      & [cycles] & (rel. \%) & [cycles] & (rel. \%) & [cycles] & (rel. \%) & speedup \\
      \cmidrule(r){1-1}
      \cmidrule(r){2-3}
      \cmidrule(r){4-5}
      \cmidrule(r){6-7}
      \cmidrule(){8-9}
      Filter            & 5\,637k    & {(15.73\%)} & 5\,637k    & {(33.96\%)} & 779k       & {(33.41\%)} & 7.24       & 4.45 \\
      Cov. matrix       & 3\,071k    & {(8.57\%)}  & 3\,071k    & {(18.50\%)} & 393k       & {(16.86\%)} & 7.81       & 10.14 \\
      Whitening         & {809k} & {(2.26\%)}  & {809k} & {(4.87\%)}  & {180k} & {(9.00\%)}  & {4.31} & {2.12} \\
      Matrix logm.      & 26\,314k   & {(73.40\%)} & 7\,070k    & {(42.58\%)} & 934k       & {(40.08\%)} & 7.60       & 2.72 \\
      \gls{svm}         & 15k        & {(0.04\%)}  & 15k        & {(0.09\%)}  & 15k        & {(0.65\%)}  & -          & 1.20 \\
      \cmidrule(r){1-1}
      \cmidrule(r){2-3}
      \cmidrule(r){4-5}
      \cmidrule(r){6-7}
      \cmidrule(){8-9}
      Total [cycles]            & {36\,912k} & & {17\,365k} & & {\textbf{2\,417k}} & & {7.18} \\
      \glspl{mac}/cycle$^\natural$ & {0.82}     & & {0.82}   & & {\textbf{5.62}} & \\
      FLOPs/cycle$^\diamond$ & 0.10     & & 0.36  & & \textbf{2.72} & \\
      instructions/cycle        & {0.845}    & & {0.870} & & {0.664} & \\
      \cmidrule(r){1-1}
      \cmidrule(r){2-3}
      \cmidrule(r){4-5}
      \cmidrule(r){6-7}
      \rebuttal{Runtime [ms]$^\star$} & \rebuttal{233.5} & & \rebuttal{115.7} & & \rebuttal{16.9} & & \\
      \rebuttal{Avg. power [mW]$^\star$} & \rebuttal{5.3} & & \rebuttal{5.2} & & \rebuttal{11.7} & & \\
      \rebuttal{Energy [\textmu J]$^\star$} & \rebuttal{1238} & & \rebuttal{602} & & \rebuttal{198} & & \\
      \bottomrule
    \end{tabular}
  \begin{tablenotes}
    \footnotesize
    \item $^\natural$ Number of fixed-point \glspl{mac} over number of cycles w/o matrix logarithms. 
    \item $^\diamond$ \acrshort{flop}/cycles during matrix logarithms, approximated by treating all floating-point instructions equally.  
    \item $^\nmid$ \glspl{mac} or \acrshort{flop} per cycle for the concurrent implementation except \gls{svm}. 
    \item $^\star$ With \gls{fc} at 50\,MHz, the cluster cores at 160\,MHz, and 0.6\,V supply. Cluster activation included (0.9\,ms, 2\,mW, 1.8\,\textmu J).
  \end{tablenotes}
    \end{threeparttable}
    }
    }
\end{table*}

\begin{figure}
  \centering
  
  \resizebox{\columnwidth}{!}{%
  \begin{tikzpicture}
    \begin{axis}[
      width=\columnwidth, height=0.6\columnwidth,
      xlabel = {\small Runtime [ms]},
      ylabel = {\small Energy [\textmu J]},
      ylabel near ticks,
      xlabel near ticks,
      xtick = {0, 7.9, 16.0, 16.9, 48.5},
      xticklabels = {0, 7.9, {16.0\hspace{1.2em}{\color{white}.}}, {\hspace{2em}16.9}, 48.5},
      ytick = {198, 210, 338, 590},
      yticklabels = {\smash{\raisebox{-0.6em}{198}}, \smash{\raisebox{0em}{210}}, 338, 590},
      xmin = 0, xmax = 75,
      ymin = 150, ymax = 650,
      legend style={
        at={(0.99, 0.79)},
        anchor=north east,
        draw=none,
        font={\footnotesize}
      },
      ]
      
      \draw[thin, black!20] (axis cs:07.9, 338.00) -- (axis cs:07.9, 0);
      \draw[thin, black!20] (axis cs:07.9, 338.00) -- (axis cs:-10, 338.00);
      \draw[thin, black!20] (axis cs:16.0, 210.39) -- (axis cs:16.0, 0);
      \draw[thin, black!20] (axis cs:16.0, 210.39) -- (axis cs:-10, 210.39);
      \draw[thin, black!20] (axis cs:48.5, 590.48) -- (axis cs:48.5, 0);
      \draw[thin, black!20] (axis cs:48.5, 590.48) -- (axis cs:-10, 590.48);
      \draw[thin, black!20] (axis cs:16.9, 198.12) -- (axis cs:16.9, 0);
      \draw[thin, black!20] (axis cs:16.9, 198.12) -- (axis cs:-10, 198.12);
      
      \addplot [thick, color=color0, mark=*] coordinates {
        (48.5, 590.479) 
        (24.9, 451.976) 
        (17.0, 402.523) 
        (13.0, 375.509) 
        (09.0, 346.994) 
        (07.9, 338.002) 
      };
      \addlegendentry{0.8\,V}
      \draw[color0] (axis cs:48.5, 590.479) -- ++(+90:0.0cm)  node[above, fill=white] {\footnotesize \color{color0} 50\,MHz};
      \draw[color0] (axis cs:24.9, 451.976) -- ++(110:0.4cm)  node[above, fill=white] {\footnotesize \color{color0} 100};
      \draw[color0] (axis cs:17.0, 402.523) -- ++(-20:0.4cm)  node[right, fill=white] {\footnotesize \color{color0} 150};
      \draw[color0] (axis cs:13.0, 375.509) -- ++(+80:0.4cm)  node[above, fill=white] {\footnotesize \color{color0} 200};
      \draw[color0] (axis cs:09.0, 346.994) -- ++(+130:0.4cm) node[above, fill=white] {\footnotesize \color{color0} 300};
      \draw[color0] (axis cs:07.9, 338.002) -- ++(-60:0.4cm)  node[below, xshift=0.1cm] {\footnotesize \color{color0} 350};
      
      \addplot [thick, color=color2, mark=*] coordinates {
        (48.0, 296.894) 
        (24.9, 236.937) 
        (16.0, 210.394)
      };
      \addlegendentry{0.6\,V}
      \draw[color2] (axis cs:48.0, 296.894) -- ++(+90:0.0cm) node[above, fill=white] {\footnotesize \color{color2} 50\,MHz};
      \draw[color2] (axis cs:24.9, 236.937) -- ++(-20:0.4cm) node[right, fill=white] {\footnotesize \color{color2} 100};
      \draw[color2] (axis cs:16.0, 210.394) -- ++(70:0.4cm) node[above]             {\footnotesize \color{color2} 160};
      
      \addplot [thick, color=color3, mark=x, mark size=2, only marks] coordinates {(16.9, 198.117)};
      \addlegendentry{Optimum}

    \end{axis}

  \end{tikzpicture}
  }
\vspace{-0.5cm}
  \caption{{Energy and runtime measurements with 0.8\,V and 0.6\,V voltage supplies, drawn respectively in blue and green, while varying the clock frequency (in MHz) of both the \gls{fc} core and the cluster cores. The energy consumption is further reduced by fine-tuning the \gls{fc} frequency to 50\,MHz while keeping the most energy-efficient configuration for the cluster cores, i.e., 160\,MHz with 0.6\,V, represented with a red cross.}}
  \label{fig:measurements}
\end{figure}
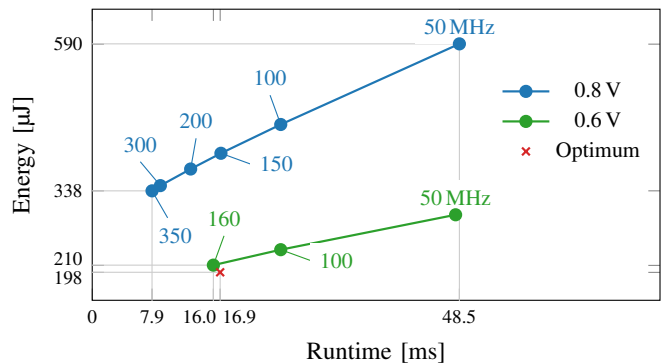

\begin{figure}[!t]
  \centering
  \resizebox{\columnwidth}{!}{%
  \begin{tikzpicture}
    \begin{axis}
    [
      no markers,
      width=\columnwidth,
      height=0.65\columnwidth,
      xlabel={\small Time [ms]},
      ylabel={\small Power [mW]},
      ylabel near ticks,
      xlabel near ticks,
      scaled ticks=false,
      xtick={0, 0.005, 0.01, 0.015, 0.02},
      xticklabels={0, 5, 10, 15, 20},
      xmin = -0.002, xmax = 0.021,
      grid=both,
    ]
      \addplot+ [name path=f, thick, color=darkgray] table [x=t, y=p, col sep=comma] {measurements/mrc_vega.csv};
      
      \draw[color0,thick,dashed] (0.0012,0) rectangle (0.0101, 15.7);
      \fill[color0,opacity=0.15] (0.0012,0) rectangle (0.00175, 15.7);
      \draw[color3,thick,dashdotted] (0.0102,0) rectangle (0.0166, 17.3);
      \fill[color3,opacity=0.15] (0.0102,0) rectangle (0.0134, 17.3);
      \draw[color2,thick,dotted, fill=color2, opacity=0.15] (0.0167,0) rectangle (0.01695, 8);
      
    \node[anchor=south] (sCL) at (axis cs: -0.0005,6){{\renewcommand{\arraystretch}{0.6}\begin{tabular}{c}\scriptsize Cluster \\ \scriptsize activ. \end{tabular}}};
    \node (dCL) at (axis cs:0.00045,3.5){};
    \draw[->](sCL)--(dCL);
    
      \draw[|{to}-{to}|] (axis cs:0.0012, 0.7) -- (axis cs:0.00175, 0.7) node [pos=0.5, anchor=south] {{\renewcommand{\arraystretch}{0.6}\begin{tabular}{c}\scriptsize 9 cores \end{tabular}}};
      \draw[{to}-{to}|] (axis cs:0.00176, 0.7) -- (axis cs:0.00230, 0.7);
      \node at (axis cs:0.00340, 0.7){\dots};
      \node at (axis cs:0.00560, 0.7){\dots};
      \node at (axis cs:0.00780, 0.7){\dots};
      \draw[|{to}-{to}|] (axis cs:0.00955, 0.7) -- (axis cs:0.0101, 0.7);
      
      \draw[|{to}-{to}|] (axis cs:0.0102,0.7) -- (axis cs:0.0134, 0.7) node [pos=0.5, anchor=south] {{\renewcommand{\arraystretch}{0.6}\begin{tabular}{c}\scriptsize 9 cores \\ \scriptsize +\,4\,\scalebox{0.9}[1]{FPUs} \end{tabular}}};
      \draw[|{to}-{to}|] (axis cs:0.0134, 0.7) -- (axis cs:0.0167, 0.7) node [pos=0.5, anchor=south] {{\renewcommand{\arraystretch}{0.6}\begin{tabular}{c}\scriptsize 9 cores \\ \scriptsize +\,4\,\scalebox{0.9}[1]{FPUs} \end{tabular}}};
    \end{axis}
  \end{tikzpicture}
  }
  \vspace{-0.5cm}
  \caption{Power measurement with the \gls{fc} at {50\,MHz} and the cluster at 160\,MHz. The colors match the compute blocks in \figref{fig:background:riemannian} explained in \secref{ch:riemann:implementation}.}
  \label{fig:results:power}
\end{figure}

We apply our methods on the BCI Competition IV-2a dataset~\cite{Brunner2008BCIA} with 22 \gls{eeg} channels, \rebuttal{i.e., $N_{ch} = 22$,} and 4 \gls{mi} classes from 9 different subjects. 
There are 288 trials for each of the training and testing sets. \new{Each trial consists of a fixation cross on the screen, followed by a cue. Afterwards the subject starts the motor imagery corresponding to the cue. The data is sampled at 250\,Hz and in this work we consider 3.5\,s time window one second after the appearance of the cue as in~\cite{Hersche2018FastFeatures}, resulting in 875 time points per \gls{eeg} channel, \rebuttal{i.e., $N_s = 875$}.}

\subsection{Accuracy, Memory Footprint, and Quantization}\label{sec:results:acc}
\gls{mrc} can be scaled to use more or fewer frequency bands and temporal windows. Hersche et al.~\cite{Hersche2018FastFeatures} have shown that $f=43$ frequency bands and a single temporal window $t=1$ can already achieve comparable accuracy (74.8\% on average) to the full \gls{mrc} (75.5\%) while requiring $3\times$ fewer features.
In this work, we use only one temporal window $t=1$ of 3.5\,s and further scale down the number of frequency bands as in~\cite{wang2021mrc}. The results show that with $2.4\times$ less frequency bands, i.e. $f=18$, of bandwidth 2\,Hz between 4 and 40\,Hz, the full precision model achieves slightly higher accuracy by introducing the regularization for the covariance matrix with the hyperparameter $\rho=1$. 
\new{
We then gain an additional 1.3\% in full precision by tuning the classifier and its hyperparameters to the subject, as shown in Table~\ref{tab:results:accuracy}. 
\rebuttal{The achieved accuracy is lower than some of the \gls{soa} works reported in Table~\ref{tab:relworks_implperf}, when considering only full precision models. Nevertheless, our \gls{mrc} requires significantly less storage for the parameters than most of the related works at the cost of minor accuracy degradation, enabling the embedded implementation on low-power \glspl{mcu}.}}

\new{\rebuttal{To efficiently deploy the proposed model, we proceed with quantization as described in \secref{ch:riemann:implementation}. This} leads to a drop in accuracy of merely 1.0\% from 75.1\% to 74.1\% when using a linear SVM for the classification step, or 1.4\% from 76.4\% to 75.0\% when selecting the best classifier for each subject as described in Section~\ref{sec:model_classifier}. The quantization allows us to make use of the SIMD extensions to improve throughput during the filtering step, the covariance matrix computation, the whitening, and finally the classification. Besides the throughput aspect, it helps keeping the memory footprint low. Our total memory requirement amounts to 84\,kB, consisting of 2$\cdot$22$\cdot$875 value for the inputs and outputs of the IIR filters in 8-bit each, 18$\cdot$(22+1)$\cdot$22/2 values for the $W_k$ in 32-bit and reused for $L_k$, 18$\cdot$(22+1)$\cdot$22/2 values for the trained feature extractor parameters $\mathbf{C}_{\text{ref,k}}^{-1/2}$ in 16-bit, and 4554$\cdot$4 for the linear SVM weights in 8-bit. 
\rebuttal{Compared to other embedded \gls{mi}-\gls{bmi} implementations, reported in Tables \ref{tab:relworks_implperf} and \ref{tab:results:accuracy}, our model achieves the highest classification accuracy.}

When using a MLP classifier, its weights need to be stored, which are larger than that of the linear SVM and dependent on its hyperparameters. This leads to a trade-off which we visualize and compare to related works in Fig.~\ref{fig:acc_mem_comp} by limiting the search space for the classifiers to those that let the system fit within the memory footprint indicated. This trade-off naturally crosses the implementation on Mr. Wolf~\cite{wang2021mrc} that was restricted to the linear SVM and expands it both to a lower as well as larger memory footprint that is only feasible on the Vega platform. It also clearly shows that our model spans the entire Pareto front within reach of typical microcontrollers as well as Mr. Wolf and Vega. The remainder of the Pareto front only includes EEG-TCNet as an additional point providing around 2\% accuracy gain at the cost of a memory footprint just slightly in excess of 1.5\,MB available on \vega. 
}

\subsection{Compute Time and Energy}

\new{Table~\ref{tab:results:riemann} shows the computation time and the performance impact of the optimizations.
Each output sample of the \gls{iir} filter requires 10 \glspl{mac}, 3 shuffle operations, and 4 bit-shifts, resulting in a theoretical maximum of 5 \glspl{mac} per cycle.
Our implementation achieves 4.45 \glspl{mac} per cycle with 7.24$\times$ parallel speedup compared to the single core implementation.
The covariance matrix computation reaches 10.14 \glspl{mac} per cycle and a parallel speedup of 7.81$\times$ using 9 cores.
The parallel speedup of the whitening is {only 4.31}$\times$ due to the parallelization overhead that is more visible with smaller matrix sizes (here 22$\times$22). On the other hand, its computation accounts for {9\%} of the whole execution time of the {final} \gls{mrc}, {meaning that any further optimizations would only yield marginal improvements}.
Contrarily, the matrix logarithm is the bottleneck of the execution, requiring {almost three quarters} of the total number of cycles in the baseline. With the improved \gls{evd}, i.e., the optimized Householder transformation, the relative runtime is reduced to {42.58}\% which translates to 3.72$\times$ speedup compared to the baseline. A further improvement is given by the parallel computation which is 7.60$\times$ and 28.17$\times$ faster compared to the single core implementation with improved \gls{evd} and the baseline, respectively.

\rebuttal{We measure the average power consumption and the execution time of all three configurations including the cluster activation and initialization, reported in Table~\ref{tab:results:riemann}. The multi-core implementation consumes 2.2$\times$ more power than the single core implementations. However, the energy consumption is 6.3$\times$ and 3$\times$ lower than the baseline and the single core implementation with improved \gls{evd}, respectively, thanks to the shorter execution time.
}

Fig.~\ref{fig:measurements} shows the energy consumption and the execution time of \gls{mrc} with linear \gls{svm} measured by varying the frequency of the \gls{fc} and the cluster cores of \vega. {In general, increasing the clock speed yields lower energy consumption, while also lowering the runtime.} With 0.6\,V voltage supply, we can clock the cores up to 160\,MHz. {This has proven to be the most energy-efficient configuration when both \gls{fc} and cluster cores run at the same frequency, consuming 210\,\textmu J including the cluster activation, the feature extraction, and the classifier. 
We further fine-tune the \gls{fc} frequency by clocking it at a lower speed, since all the computation is delegated to the cluster. By reducing it down to 50\,MHz, the energy consumption is further reduced to 198\,\textmu J, represented with a red cross in Fig.~\ref{fig:measurements}.} 
Compared to the previous \gls{soa} low-energy embedded implementation for \gls{mi}-\gls{bmi}, i.e., Q-EEGNet~\cite{Schneider2020}, it consumes {1.7$\times$} less energy while being 3.2\% more accurate. The accuracy can be further increased by 0.9\% when subject-specific classifiers are trained, as discussed in Sec.~\ref{sec:results:acc}. Comparing to the \gls{mrc} on \wolfe, it is {6.6$\times$} more energy-efficient and requires less computation time (16.90\,ms vs. 33.39\,ms~\cite{wang2021mrc}). The runtime can be further shortened by clocking the cores at higher frequency. With the supply voltage of 0.8\,V, and both domains clocked at 350\,MHz, we obtain an energy consumption of {338\,\textmu J} and a runtime of only {7.9\,ms}.

Fig.~\ref{fig:results:power} depicts the measured power trace in the most energy-efficient configuration.
As explained in Sec.~\ref{ch:riemann:implementation}, the filtering, the covariace matrix, and the whitening are executed consecutively, framed with blue dashed line, while each step is computed in parallel using all cluster cores, colored with blue background. This is clearly visible from the 18 repeated patterns in the measure power trace.
An additional pattern follows repeated twice. It reflects the parallel computation of the matrix logarithm, framed with red dashdotted line. The 18 matrix logarithms are computed, distributed to the 9 cores on a first-come first-served schedule. Unlike on \wolfe, where this workload is \rebuttal{unbalanced} for its 8 parallel cores and contributes negatively to the parallel speedup, the work can be evenly distributed to the 9 cores of \vega with an optimal execution. 
Moreover, with 4 \glspl{fpu} instead of 2, our implementation reaches 2.72 \glspl{flop} per cycle compared to the 1.69 of~\cite{wang2021mrc}. The ideal \glspl{flop}/cycle is not reached due to the divisions and the square root operations.}
Finally, the \gls{svm} accounts for a minimal part of the execution with 0.15\,ms. 

%

\new{
\section{Conclusion}\label{ch:conclusion}
This paper presents an improved \gls{mrc} with a reduced model size while keeping comparable accuracy (75.1--76.4\% vs. 75.5\%~\cite{Hersche2018FastFeatures}), allowing accurate low-power \rebuttal{embedded} \gls{bmi}. We further scale down the model by quantizing and proposing a mixed-precision implementation yielding a minimal accuracy loss of 1.4\%, which is still 3.7\% more accurate than the \gls{soa} embedded \gls{cnn} for \gls{bmi} named Q-\eegnet{}~\cite{Schneider2020}. We propose a parallel implementation on a low-power \gls{mcu} called \vega, which takes only {7.9}\,ms and consumes {338}\,$\mu$J, or alternatively takes {16.9}\,ms using merely {198}\,$\mu$J to classify the 3.5s time window of data---{57$\mu$W} when continuously acquiring new data and processing it immediately, or {85$\mu$W} when overlapping the 3.5s samples by 50\% to avoid missing any user inputs.
We provide an insight on accuracy-cost trade-off for embedded \gls{bmi} models with actual implementation and measurements and claim highest accuracy implementation of EEG MI-BMI of any current work within the memory constraints of current MCUs while consuming the least energy. 
}



\bibliographystyle{IEEEtran}
\bibliography{bib,bstctl}

\begin{IEEEbiography}[{\includegraphics[width=1in,height=1.25in,clip,keepaspectratio]{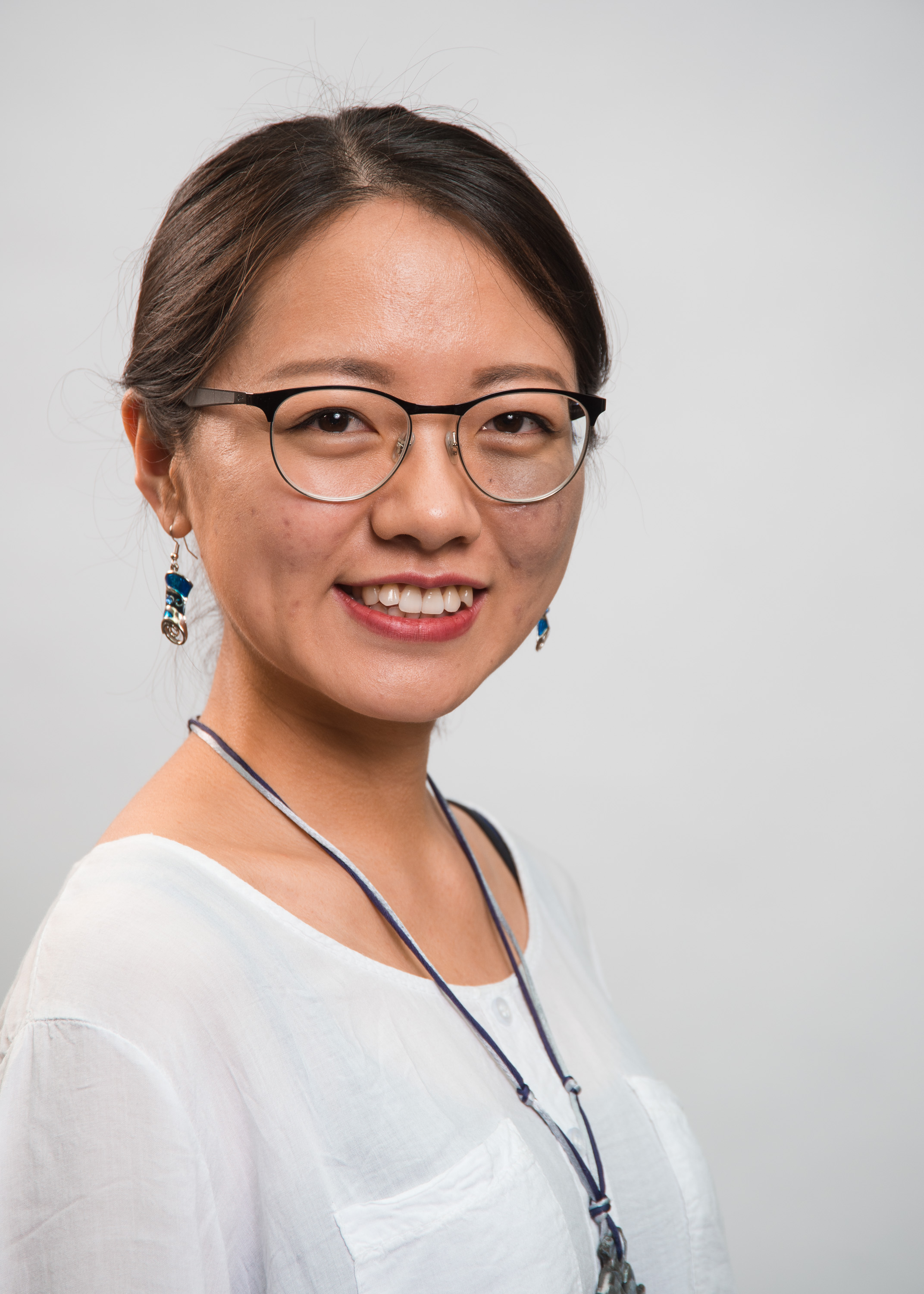}}]{Xiaying Wang}
received her B.Sc. and M.Sc. degrees in biomedical engineering from Politecnico di Milano, Italy and ETH Zürich, Switzerland in 2016 and 2018, respectively. She is currently pursuing a Ph.D. degree at the Integrated Systems Laboratory at ETH Zürich. Her research interests include biosignal processing, brain--machine interface, low power embedded systems, energy-efficient smart sensors, and machine learning on microcontrollers. She received the excellent paper award at the IEEE Healthcom conference in 2018 and she won the Ph.D. Fellowship funded by the Swiss Data Science Center in 2019. \vspace{-0.3cm}
\end{IEEEbiography}

\begin{IEEEbiography}[{\includegraphics[width=1in,height=1.25in,clip,keepaspectratio]{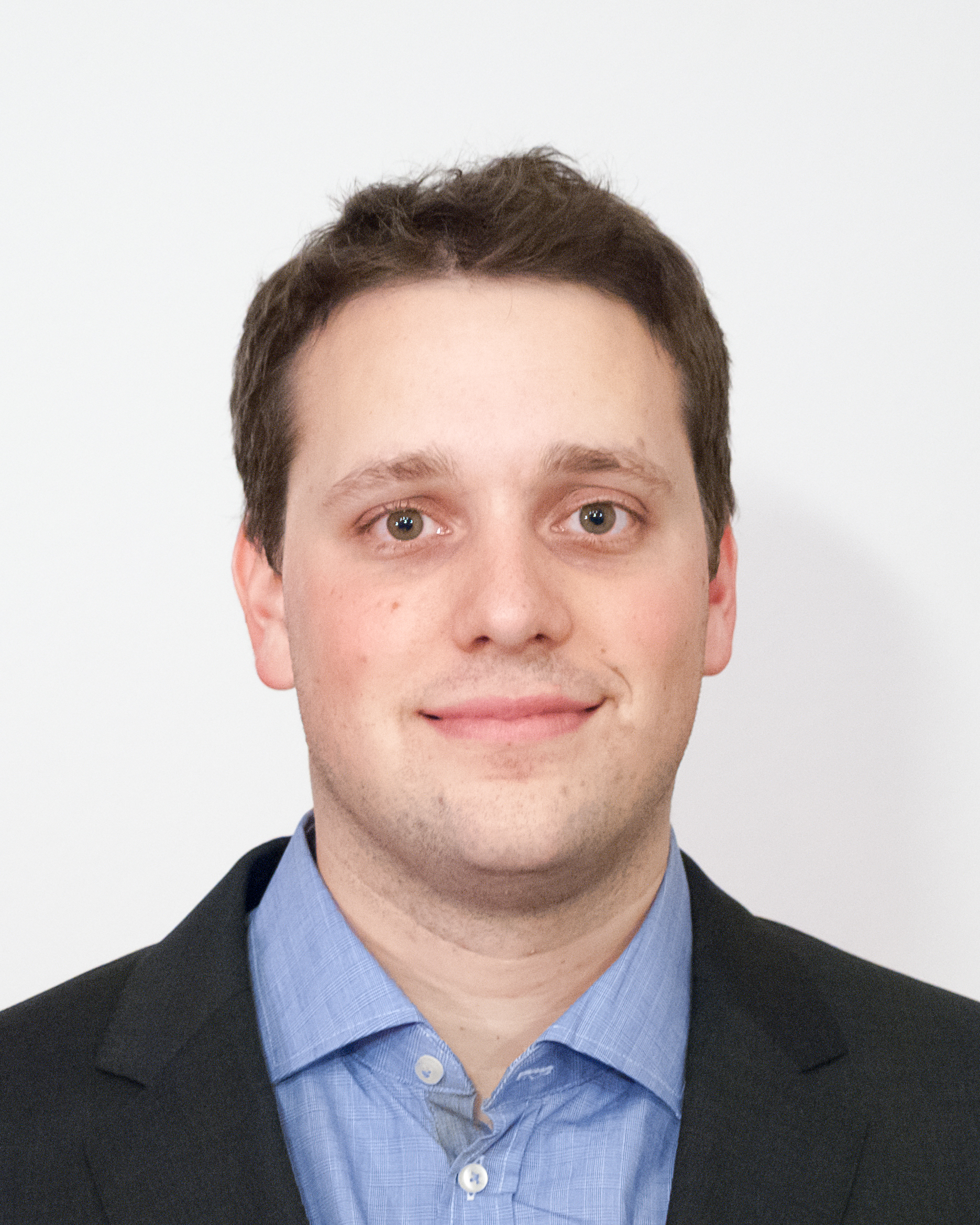}}]{Lukas Cavigelli}
received the B.Sc., M.Sc., and Ph.D. degree in electrical engineering and information technology from ETH Zürich, Zürich, Switzerland in 2012, 2014 and 2019, respectively. After spending a year as a Postdoc at ETH Zürich, he has joined Huawei's Zurich Research Center in Spring 2020. His research interests include deep learning, computer vision, embedded systems, and low-power integrated circuit design. He has received the best paper award at the VLSI-SoC and the ICDSC conferences in 2013 and 2017, the best student paper award at the Security+Defense conference in 2016, and the Donald O. Pederson best paper award (IEEE TCAD) in 2019. \vspace{-0.3cm}
\end{IEEEbiography}

\begin{IEEEbiography}[{\includegraphics[width=1in,height=1.25in,clip,keepaspectratio]{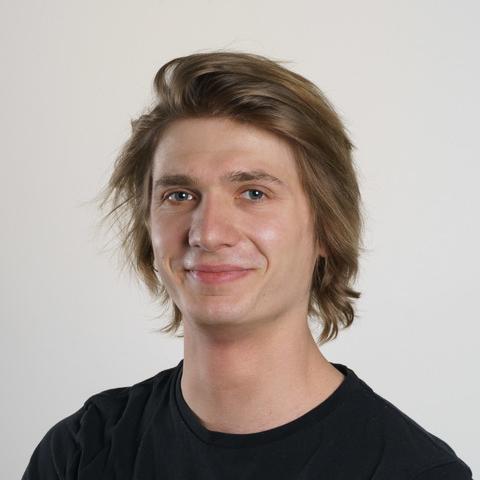}}]{Tibor Schneider}
received his B.Sc and M.Sc in electrical engineering and
Information Technology at Hochschule für Technik Rapperswil and ETH
Zurich, Switzerland in 2017 and 2021, respectively. Since 2021, he is a
Ph.D student at the Networked Systems Group at ETH Zürich, advised by
Laurent Vanbever. His current research is aimed at improving
verification and synthesis methods for configurations of internet
networks. In 2021, he won the ABB Research Prize at ETH Zürich. \vspace{-0.3cm}
\end{IEEEbiography}

\begin{IEEEbiography}[{\includegraphics[width=1in,height=1.25in,clip,keepaspectratio]{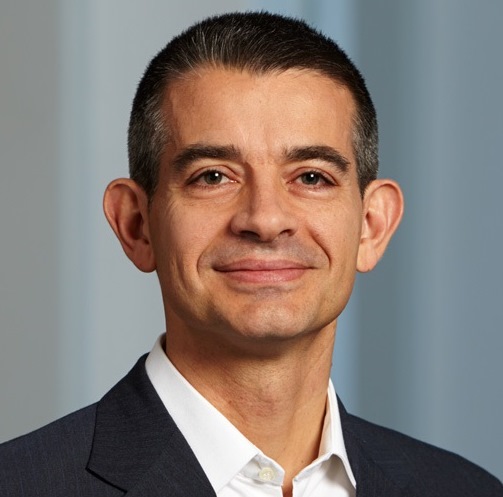}}]{Luca Benini}
is the Chair of Digital Circuits and Systems at ETH Zürich and a Full Professor at the University of Bologna. He has served as Chief Architect for the Platform2012 in STMicroelectronics, Grenoble. Dr. Benini’s research interests are in energy-efficient system and multi-core SoC design. He is also active in the area of energy-efficient smart sensors and sensor networks. He has published more than 1’000 papers in peer-reviewed international journals and conferences, four books and several book chapters. He is a Fellow of the ACM and of the IEEE and a member of the Academia Europaea. \vspace{-0.3cm}
\end{IEEEbiography}

\end{document}